\documentclass[nofootinbib,preprint,superscriptaddress,showpacs,preprintnumbers,amsmath,amssymb,aps,pre]{revtex4-2}
\usepackage[utf8]{inputenc} 
\usepackage{footnote}
\usepackage{epsfig}
\usepackage{natbib}
\usepackage{booktabs}
\usepackage{amsmath}
\usepackage{times}
\usepackage{epstopdf}
\usepackage{graphicx}
\usepackage{color}

\usepackage{afterpage}

\begin{document}

\title{Period-doubling bifurcations and islets of stability in two-degree-of-freedom Hamiltonian systems}

\author{Alexandre R. Nieto}
\email[]{alexandre.rodriguez@urjc.es}
\affiliation{Nonlinear Dynamics, Chaos and Complex Systems Group, Departamento de
F\'{i}sica, Universidad Rey Juan Carlos, Tulip\'{a}n s/n, 28933 M\'{o}stoles, Madrid, Spain}

\author{Jes\'{u}s M. Seoane}
\affiliation{Nonlinear Dynamics, Chaos and Complex Systems Group, Departamento de
F\'{i}sica, Universidad Rey Juan Carlos, Tulip\'{a}n s/n, 28933 M\'{o}stoles, Madrid, Spain}

\author{Miguel A.F. Sanju\'{a}n}
\affiliation{Nonlinear Dynamics, Chaos and Complex Systems Group, Departamento de
F\'{i}sica, Universidad Rey Juan Carlos, Tulip\'{a}n s/n, 28933 M\'{o}stoles, Madrid, Spain}
\affiliation{Department of Applied Informatics, Kaunas University of Technology, Studentu 50-415, Kaunas LT-51368, Lithuania}

\date{\today}

\begin{abstract}

In this paper, we show that the destruction of the main KAM islands in two-degree-of-freedom Hamiltonian systems occurs through a cascade of period-doubling bifurcations. We calculate the corresponding Feigenbaum constant and the accumulation point of the period-doubling sequence. By means of a systematic grid search on exit basin diagrams, we find the existence of numerous very small KAM islands (\textquotedblleft islets") for values below and above the aforementioned accumulation point. We study the bifurcations involving the formation of islets and we classify them in three different types. Finally, we show that the same types of islets appear in generic two-degree-of-freedom Hamiltonian systems and in area-preserving maps.

\end{abstract}

\pacs{05.45.Ac,05.45.Df,05.45.Pq}
\maketitle
\newpage

\section{Introduction} \label{sec:Introduction}

One of the most remarkable characteristics of conservative nonlinear systems, such as area-preserving maps and non-integrable Hamiltonians, is the existence of Kolmogorov-Arnold-Moser (KAM) tori surrounding stable periodic orbits. Embedded in a chaotic sea, KAM tori constitute regions (\textquotedblleft islands") of stability where periodic and quasiperiodic motions take place. Nonetheless, the inner structure of KAM islands is anything but simple. As shown by the Poincaré-Birkhoff theorem \cite{Poincare,Birkhoff}, resonant islands are constantly created around the main stable periodic orbit. Near these resonant islands, chaotic orbits can exist and form an inner chaotic domain \cite{Greene79}. As a result, chaotic and regular trajectories coexist within KAM islands, and they are separated from the chaotic sea by a boundary known as the \textquotedblleft last KAM curve" \cite{Contopoulos99}.

 As the parameters of the system are modified, the structure of the KAM islands evolves in a complex manner. Even though the presence of KAM islands is directly explained by the existence of stable periodic orbits, they undergo an infinite set of bifurcations that generate a fractal tree-like structure that has been firstly shown in a paper by Greene \textit{et al.} \cite{Greene81}. The ramifications appearing in the top of these structures are a consequence of a sequence of period-doubling bifurcations similar to the ones studied by Feigenbaum in the case of dissipative systems \cite{Feigenbaum}. This analogous behavior observed in both dissipative and conservative systems lead to intensive efforts to numerically characterize the sequences of period-doubling bifurcations in conservative systems. So much so that during the early '80s of the past century, within only a few years different authors obtained that in two-dimensional area-preserving maps the Feigenbaum constant takes the value $\delta_H\approx8.721$ \cite{Benettin80a,Collet,Bountis} (we recall that the dissipative Feigenbaum constant is $\delta\approx4.669$). Some years later, these results have been extended to four-dimensional volume-preserving maps \cite{Mao85}.
  
In the case of continuous-time Hamiltonian systems, the literature is filled with countless articles studying periodic orbits and their close relation with KAM tori. Some early works are \cite{Contopoulos71,Contopoulos83,Aguiar87,Mao92,Contopoulos96}, while more recent research can be found in \cite{Manchein13,Barrio20,Nieto22a}. Undoubtedly, one of the disadvantages of Hamiltonian systems when compared with discrete ones is the computational cost of the numerical simulations and, in this context, the difficulty to accurately detect periodic orbits. As a consequence, numerous research works have focused the attention on developing new methods and techniques to search for periodic orbits \cite{Helleman,Hadji06,Barrio09,Abad11}. Nonetheless, despite the wide variety of techniques for computing periodic orbits, the period-doubling cascades have not been exhaustively explored in two-degree-of-freedom Hamiltonian systems and, as far as we know, the conservative Feigenbaum constant has not been obtained in this kind of systems. In this paper, we use a two-degree-of freedom-Hamiltonian system to describe the destruction of the main tori in terms of the period-doubling cascade. We also calculate the conservative Feigenbaum constant, obtaining the same value that was found in discrete conservative systems, as indicated above. 

Based on previous research, one might assume that the structure and evolution of KAM islands can be fully understood by studying the bifurcations of the main stable periodic orbit. Additionally, by numerically obtaining the accumulation point (also known as Feigenbaum point) of the period-doubling sequence, the exact parameter value at which the last KAM tori are destroyed can be determined. Over this value, the reign of chaos begins. However, research conducted in the '80s of the past century discovered that typical area-preserving maps exhibit very small KAM islands (\textquotedblleft islets") even for parameter values significantly above the accumulation point \cite{Mackay82}. This finding was corroborated years later by Contopoulos \textit{et al.}, who found that these islets of stability were not related to the main tori, but instead seemed to appear in saddle-node bifurcations out in the chaotic sea. Recently, islets of stability have also been found in two-degree-of-freedom Hamiltonian systems \cite{Barrio09NJP}. Moreover, it has been demonstrated through computer-assisted proofs that they are not a product of spurious numerical simulations \cite{Barrio20}.

 Although islets occupy a small volume in phase space and appear in a reduced range of parameter values, their existence implies that the system dynamics is not fully governed by chaos. Moreover, even small KAM islands can influence nearby chaotic trajectories through their stickiness \cite{Contopoulos10,Altmann06}, as well as affect global system properties such as transport \cite{Mackay84,Zaslasvky02} and decay correlations \cite{Karney83}. In this manuscript, we have conducted a comprehensive search for islets and we have found many of them below and above the accumulation point. After carefully analyzing the bifurcations involved in their formation, we have classified them into three different types.

The manuscript is organized as follows. First, in Sec.~\ref{sec2}, we introduce the model used in this work and the methods for computing periodic orbits and their stability. The description of the destruction of the main tori, together with the numerical computation of the conservative Feigenbaum constant is shown in Sec.~\ref{sec3}. The analysis and classification of islets is carried out in Sec.~\ref{sec4}. To illustrate the generality of the previous results, in Sec.~\ref{sec5} we show that the same types of islets also appear in different Hamiltonian systems and even in the case of area-preserving maps. Finally, in Sec.~\ref{sec6}, we present the main conclusions of this manuscript.

\section{Model description}\label{sec2}

For this research, we chose the Hénon-Heiles system \cite{HH64} as our model. This system is a well-known example of a two-degree-of-freedom Hamiltonian and has been extensively studied in the field of nonlinear dynamics. It was named after the French astronomer Michel Hénon and the American astrophysicist Carl Heiles, who used it in 1964 to search for the third integral of motion. The Hamiltonian describing this system is given by:
\begin{equation}
	{\cal{H}}=\frac{1}{2}(\dot{x}^2+{\dot{y}}^2)+\frac{1}{2}(x^2+y^2)+x^2y-\frac{1}{3}y^3.
\end{equation}

As a consequence, the equations of motion read:

\begin{equation} \label{eq_motion}
	\begin{aligned}
		\dot{x} & = p_x, \\
		\dot{y} & = p_y, \\
		\dot{p_x} & = -x - 2xy , \\
		\dot{p_y} &= -y - x^2 + y^2.
	\end{aligned}
\end{equation}

\begin{figure}[h!]
	\centering
	\includegraphics[clip,height=7.4cm,trim=0cm 0cm 0cm 0cm]{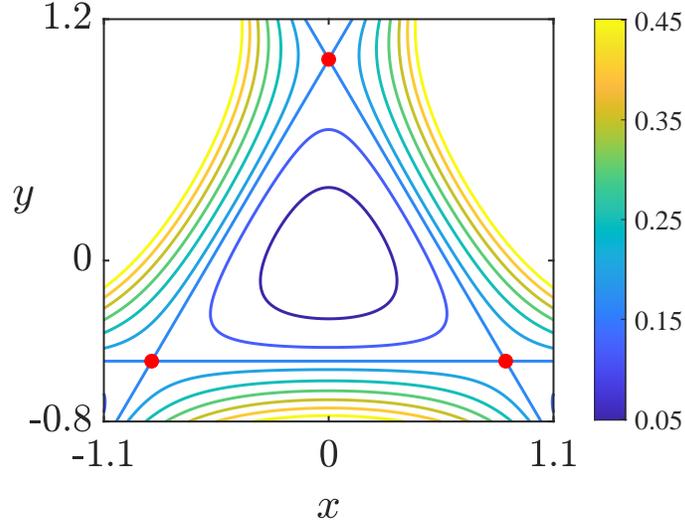}
	\caption{Isopotential curves of the Hénon–Heiles system for different values of the
		potential $V(x, y) =\frac{1}{2}(x^2+y^2)+x^2y-\frac{1}{3}y^3.$ The curves are color-coded based on the value of the potential, as indicated by the accompanying color bar. Values below and above the escape energy $E_e = 1/6$ are displayed. The three saddle points of the potential are indicated on the plot by red dots.}
	\label{Fig1}
\end{figure}

Since the Hamiltonian function governing the Hénon-Heiles system has no time dependence, the energy is conserved and can be expressed as ${\cal{H}}(x,y,p_x,p_y)=E$. Above the threshold $E_e=1/6$, known as escape energy, the potential exhibits three symmetric exits separated by an angle of $2\pi/3$ radians, as can be seen in Fig.~\ref{Fig1}. When the energy exceeds $E_e$, the particles can escape towards $\pm\infty$ through one of these exits. Conversely, when the energy is below $E_e$, the motion of the particles is bounded.

\begin{figure}[h!]
	\centering
	\includegraphics[clip,height=7.2cm,trim=0cm 0cm 0cm 0cm]{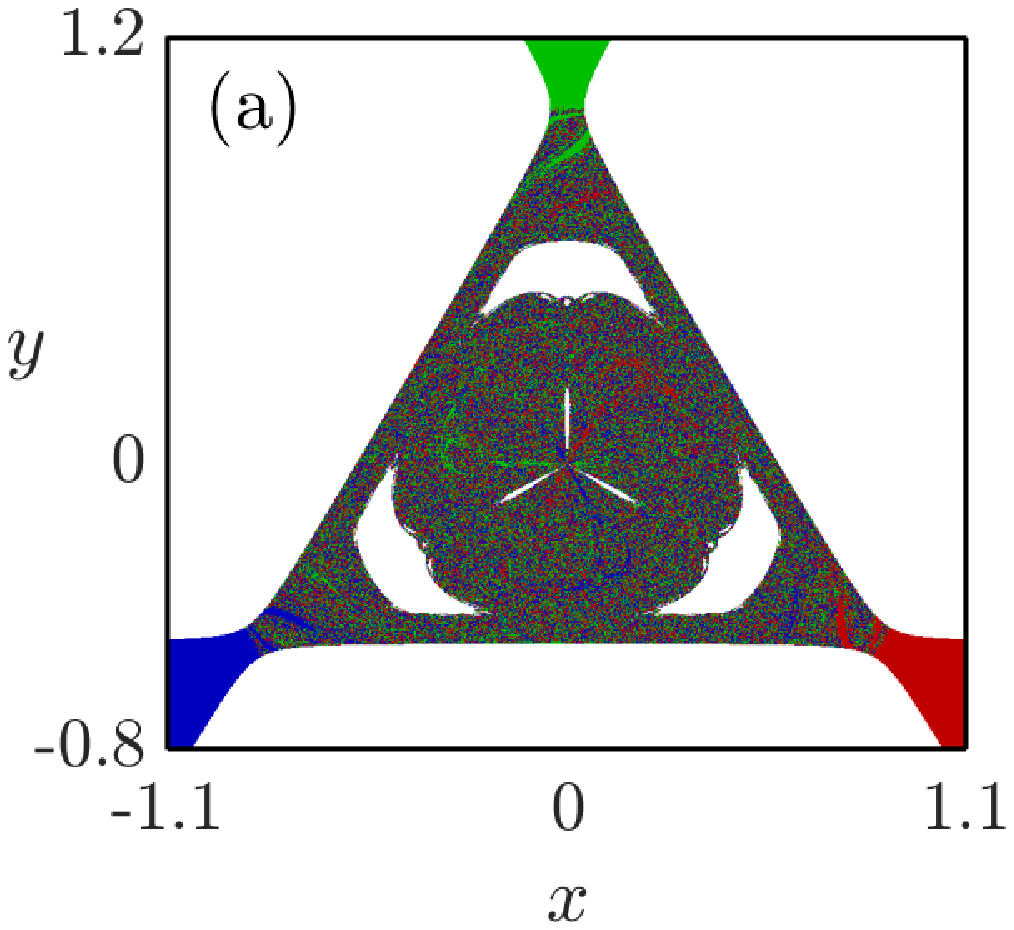}
	\includegraphics[clip,height=7.2cm,trim=0cm 0cm 0cm 0cm]{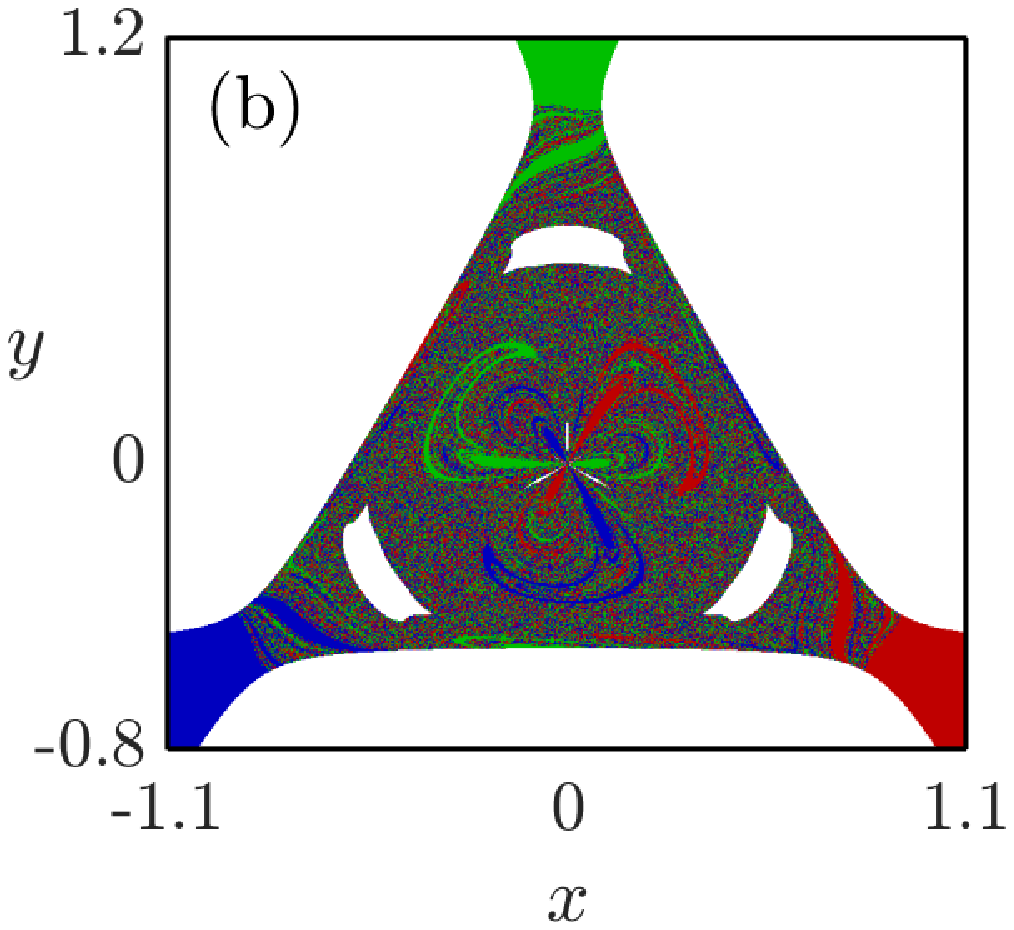}
	\caption{Exit basins in the physical space of the Hénon-Heiles system with energy (a) $E=0.17$ and (b) $E=0.18$. The colors red, green and blue refer to initial conditions leading to the three exits of the potential: Exit $1$ ($y\to\infty$), Exit $2$ ($x,y\to-\infty$), and Exit $3$ ($x\to\infty,y\to-\infty$). White regions inside the potential correspond to KAM islands.}
	\label{Fig2}
\end{figure}

The fact that the Hénon-Heiles system exhibits escapes allows us to define exit basins \cite{Contopoulos02,Aguirre01}. Similarly to basins of attraction in dissipative systems, exit basins are sets of initial conditions that lead to escape through a specific exit of the potential. Since initial conditions within a KAM island do not escape, it is possible to accurately detect the external structure of KAM islands by computing exit basin diagrams. This approach reduces computational cost compared to closed systems, where a systematic search for KAM islands requires the use of chaos indicators such as SALI or GALI \cite{Skokos04,Skokos07}. As an example, we show exit basin diagrams for two values of the energy ($E=0.17$ and $E=0.18$) in Fig.~\ref{Fig2}. The colors green, red, and blue indicate initial conditions escaping through exits $1$ ($y\to\infty$), $2$ ($x,y\to-\infty$), and $3$ ($x\to\infty,y\to-\infty$), respectively. The white regions inside the potential correspond to initial conditions that never escape, so they constitute KAM islands. \\

Using a simple tool like the exit basin diagrams, we can find KAM islands and detect with high accuracy their external structure. Hence, for a complete description of their evolution and destruction we only need to compute the associated periodic orbits and their stability.

The Hénon-Heiles system, like most Hamiltonian systems, has some symmetries. In particular, the system is time-reversible and possesses the symmetry group of an equilateral triangle ($D_3$ symmetry). As a consequence, its periodic orbits are also symmetric. They can be symmetric with respect to the three symmetry axes or only with respect to one of them. On the latter case, there necessarily exist two additional periodic orbits that are symmetric with respect to the other two symmetry axes. Due to these symmetry arguments, all periodic orbits must perpendicularly cross one of the three symmetry axes. For convenience, we find periodic orbits that are symmetric about the $y$-axis. Hence, any trajectory that starts at $x_0=0$ being perpendicular to the $y$-axis (i.e., $\dot{y}_0=0$ and $\dot{x}_0=f(y_0,E)$) and that eventually crosses perpendicularly again the same axis corresponds to a periodic orbit. The number of crossings between perpendicular intersections is the multiplicity $m$ of the periodic orbit. On the other hand, the period $T$ of a periodic orbit is twice the time needed to return perpendicularly to the $y$-axis. Therefore, the condition for a periodic orbit to exist is $x(0,y_0,\dot{x}_0,0;T/2)=\dot{y}(0,y_0,\dot{x}_0,0;T/2)=0$.

Consequently, we have computed periodic orbits following the systematic search for symmetric periodic orbits described in \cite{Barrio09}. We have determined the stability of periodic orbits by means of the eigenvalues of the monodromy matrix $M(T)$, which is the solution at time $T$ (one period of the orbit) of the linear matrix differential system
\begin{equation}
	\dot{M}= 
	\begin{pmatrix}
		0 & I_2 \\
		-\mbox{Hess}(V(x,y)) & 0	 
	\end{pmatrix}M \kern 2pc \mbox{with} \kern 0.2pc M(0)=I_4,
\end{equation}
being $\mbox{Hess}(V(x,y))$ the Hessian matrix of the potential function and $I_n$ denotes the identity matrix of order $n$.

Since $M(T)$ is a real symplectic matrix, its eigenvalues need not be explicitly calculated. Instead, the stability can be determined using the stability index $\kappa=\mbox{tr}(M(T))-2$ \cite{Henon69}. In particular, a periodic orbit is stable if $|\kappa|<2$, unstable if $|\kappa|>2$, and critical if $|\kappa|=2$.

\newpage
\section{The destruction of the main KAM island}\label{sec3}
The Hénon-Heiles system features a main KAM island that surrounds a stable periodic orbit and its bifurcation branches. The bifurcations that occur in the branches of periodic orbits before they become unstable have been profoundly studied in \cite{Barrio20,Nieto22a}. Here, we focus our attention on the period-doubling bifurcations that destroy the main family of periodic orbits and cause the main KAM island to disappear.

\begin{figure}[h!]
	\centering
	\includegraphics[clip,height=6.1cm,trim=0cm 0cm 0cm 0cm]{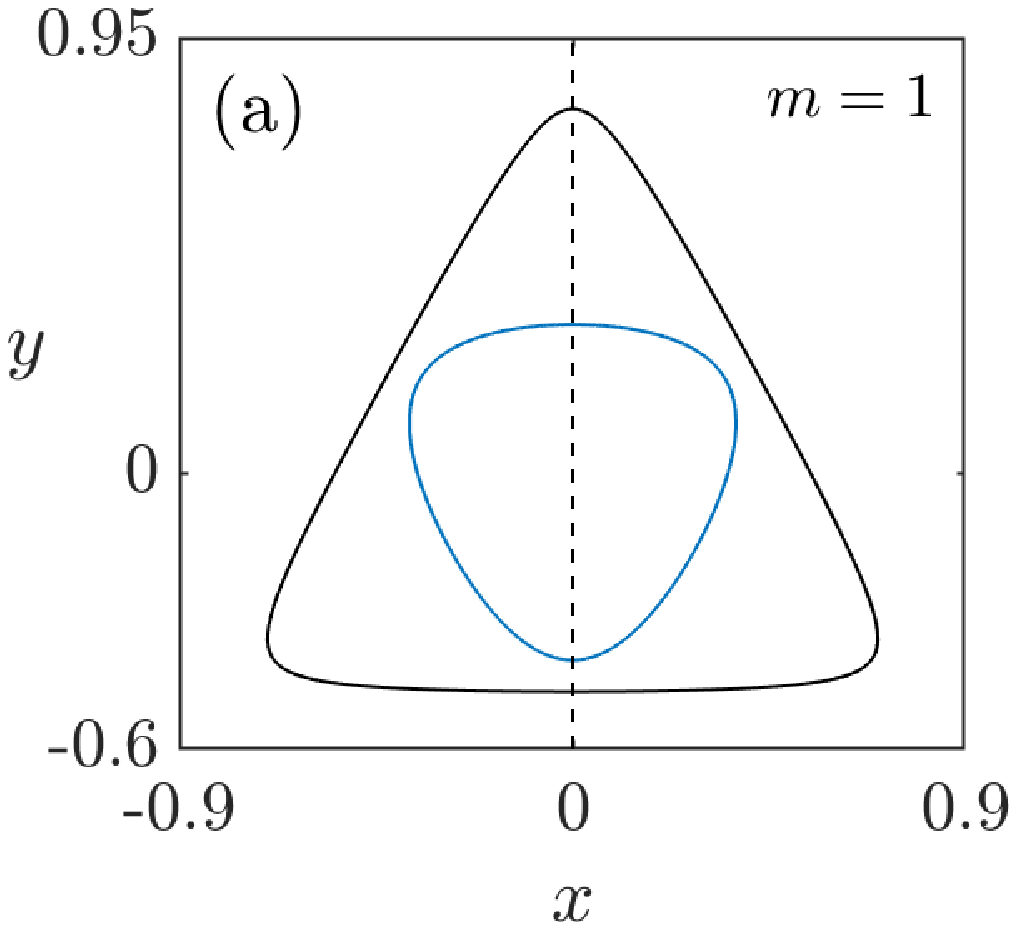}
	\includegraphics[clip,height=6.1cm,trim=0cm 0cm 0cm 0cm]{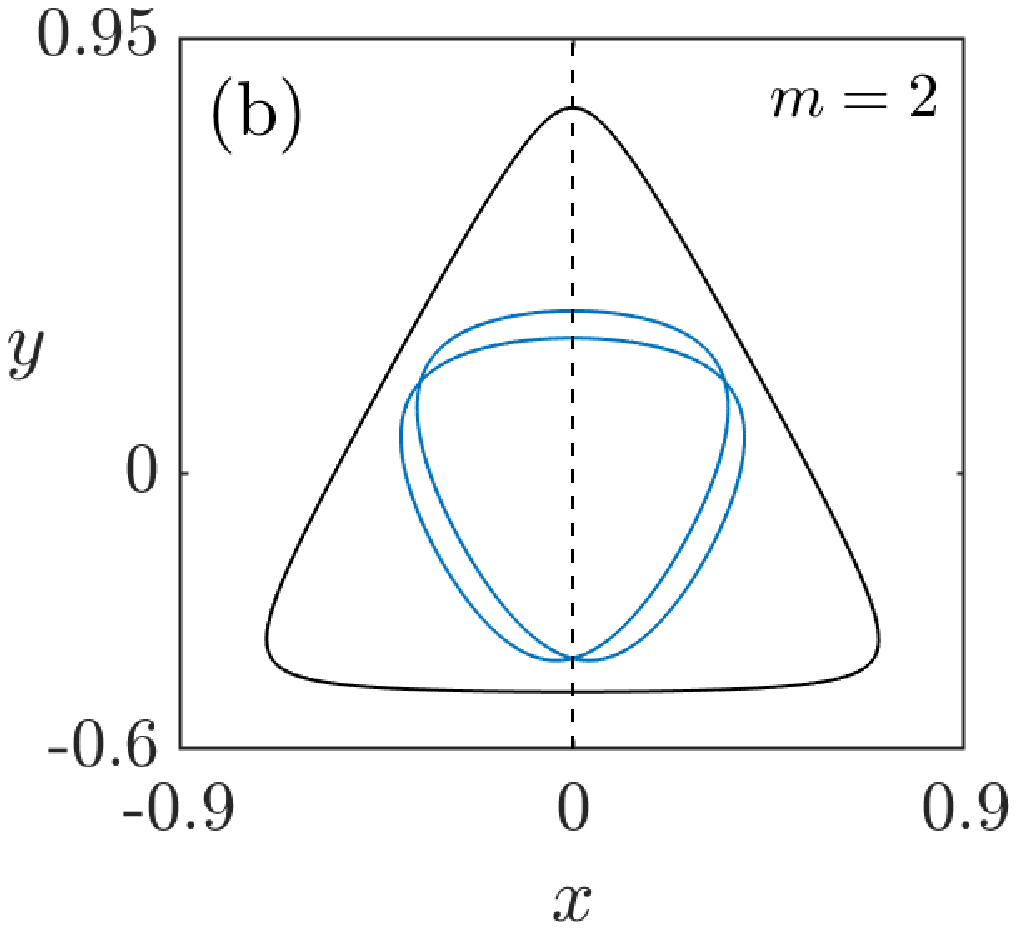}
	\includegraphics[clip,height=6.1cm,trim=0cm 0cm 0cm 0cm]{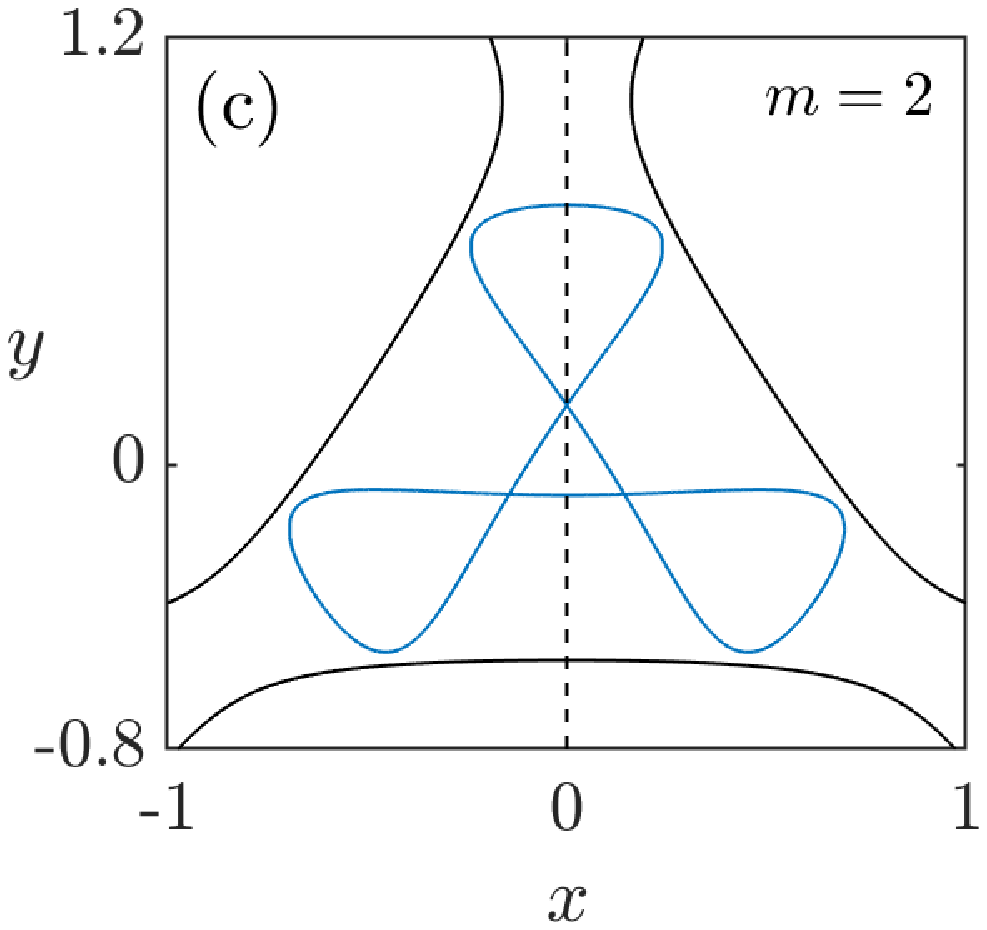}
	\includegraphics[clip,height=6.1cm,trim=0cm 0cm 0cm 0cm]{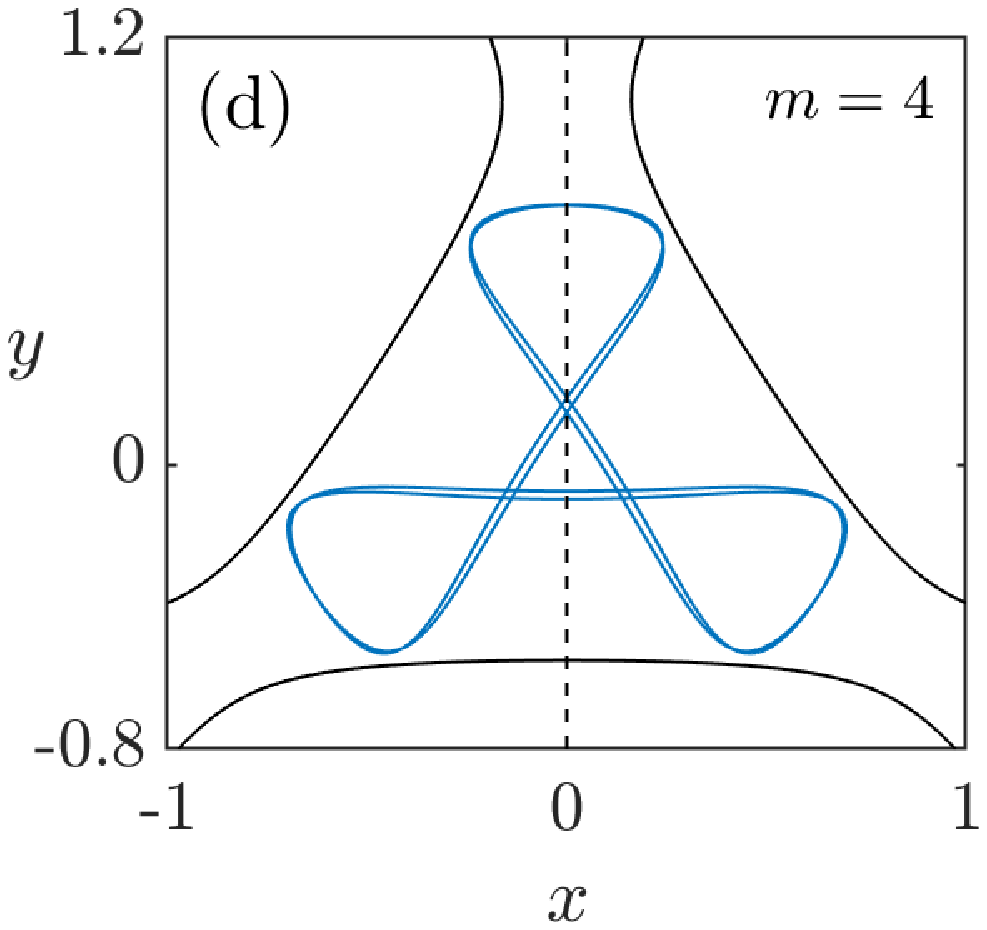}
	\caption{Periodic orbits in the Hénon-Heiles system for energy values (a) $E=0.1486$, (b) $E=0.1488$, (c) $E=0.2062$, and (d) $E=0.2064$. The multiplicity $m$ of the orbits is indicated in each panel. Orbits depicted in panels (a-b) and (c-d) have been computed for energy values just prior to and immediately following the first and second period-doubling bifurcations, respectively.}
	\label{Fig3}
\end{figure}

For low energy values, the main KAM island surrounds a periodic orbit of multiplicity $m=1$. For energies near zero, the periodic orbit takes on an almost circular shape due to the system behaving like a two-dimensional harmonic oscillator. At higher energies, the orbit exhibits a triangular symmetry, as shown in Fig.~\ref{Fig3}(a) for $E=0.1486$. By slightly increasing the energy until $E_1\approx0.14865$, the periodic orbit loses its stability and a stable periodic orbit of double multiplicity emerges (see Fig.~\ref{Fig3}(b)). Therefore, the first period-doubling bifurcation has occurred. Further increasing the energy causes the shape of the $m=2$ periodic orbit to evolve until becoming almost unrecognizable, as illustrated in Fig.~\ref{Fig3}(c). Following the same fate of its parent periodic orbit, this $m=2$ periodic orbit loses its stability in the subsequent period-doubling bifurcation, which occurs for $E_2\approx0.20626$. The newly bifurcated $m=4$ periodic orbit is depicted in Fig.~\ref{Fig3}(d). This sequence of period-doubling bifurcations continues until reaching the accumulation point $E_\infty$, where the last bifurcation branches become unstable. As a consequence, beyond $E_\infty$ large KAM islands do not exist anymore in the system.

\begin{figure}[h!]
	\centering
	\includegraphics[clip,height=6.4cm,trim=0cm 0cm 0cm 0cm]{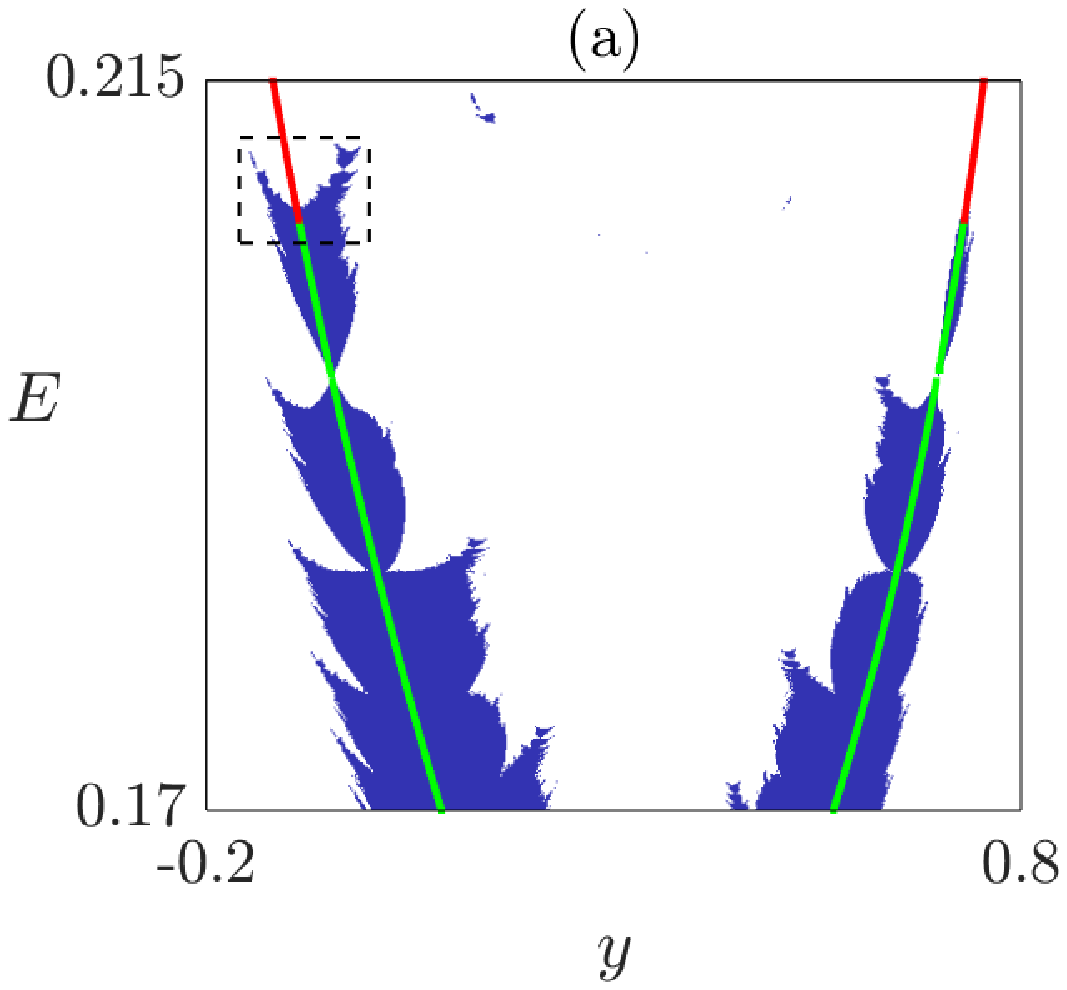}
	\includegraphics[clip,height=6.4cm,trim=0cm 0cm 0cm 0cm]{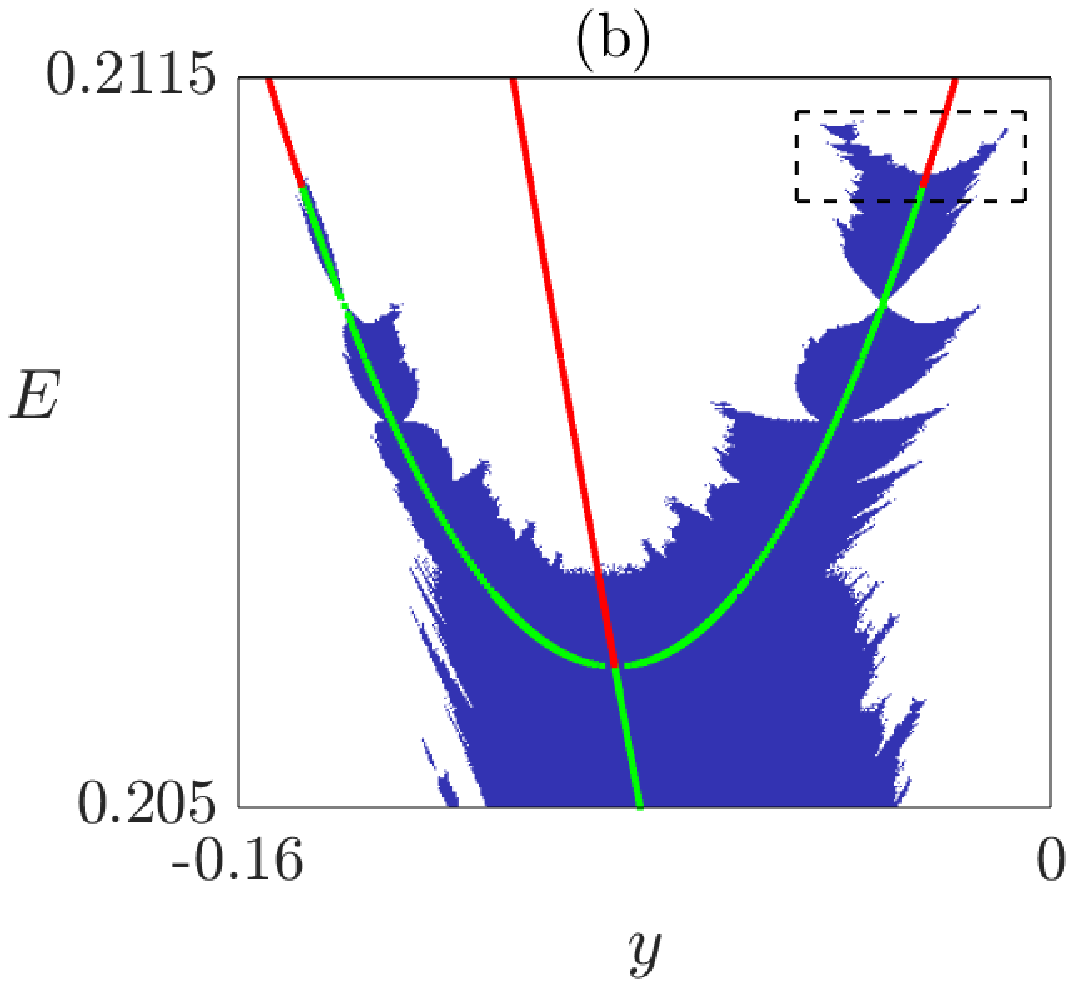}\\\vspace{0.4cm}
	\includegraphics[clip,height=6.4cm,trim=0cm 0cm 0cm 0cm]{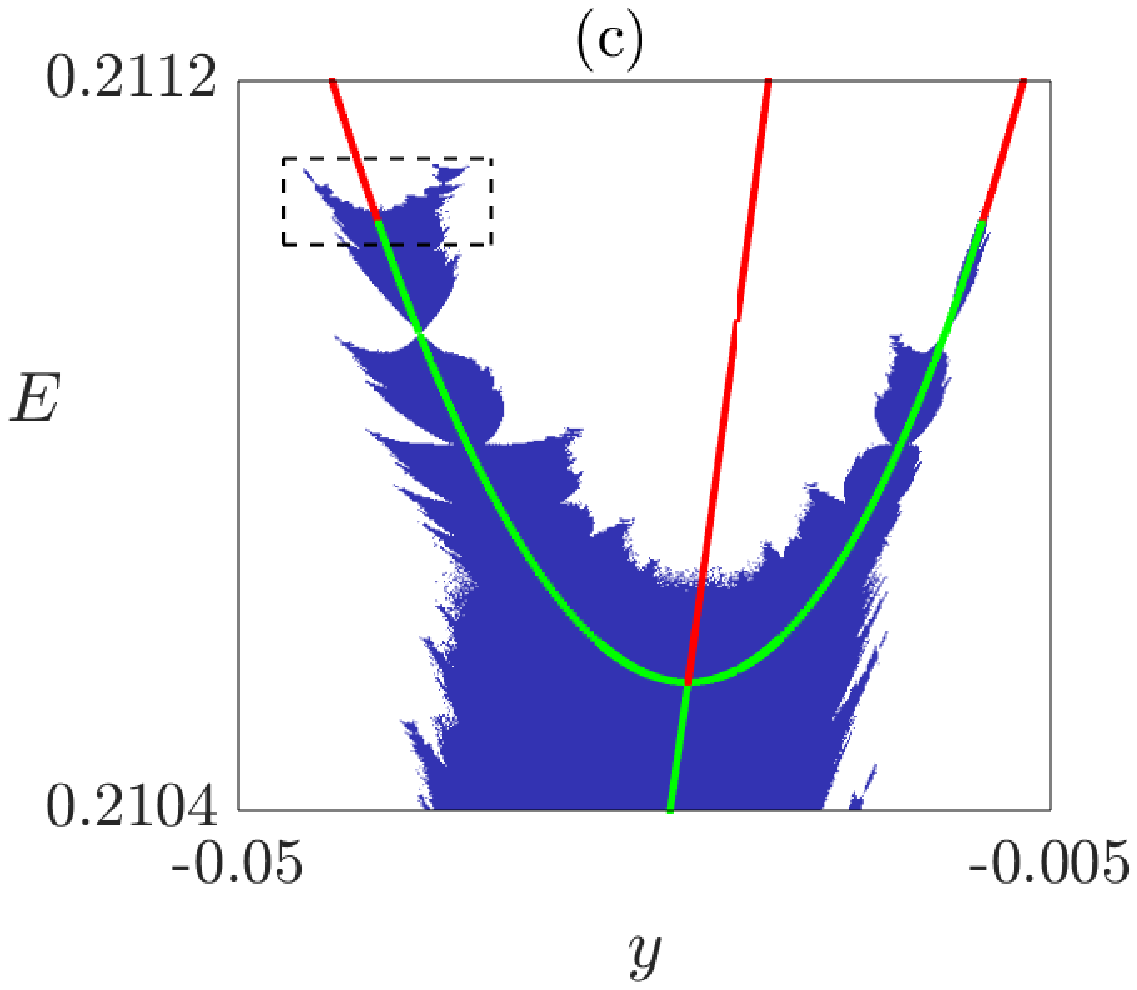}
	\includegraphics[clip,height=6.4cm,trim=0cm 0cm 0cm 0cm]{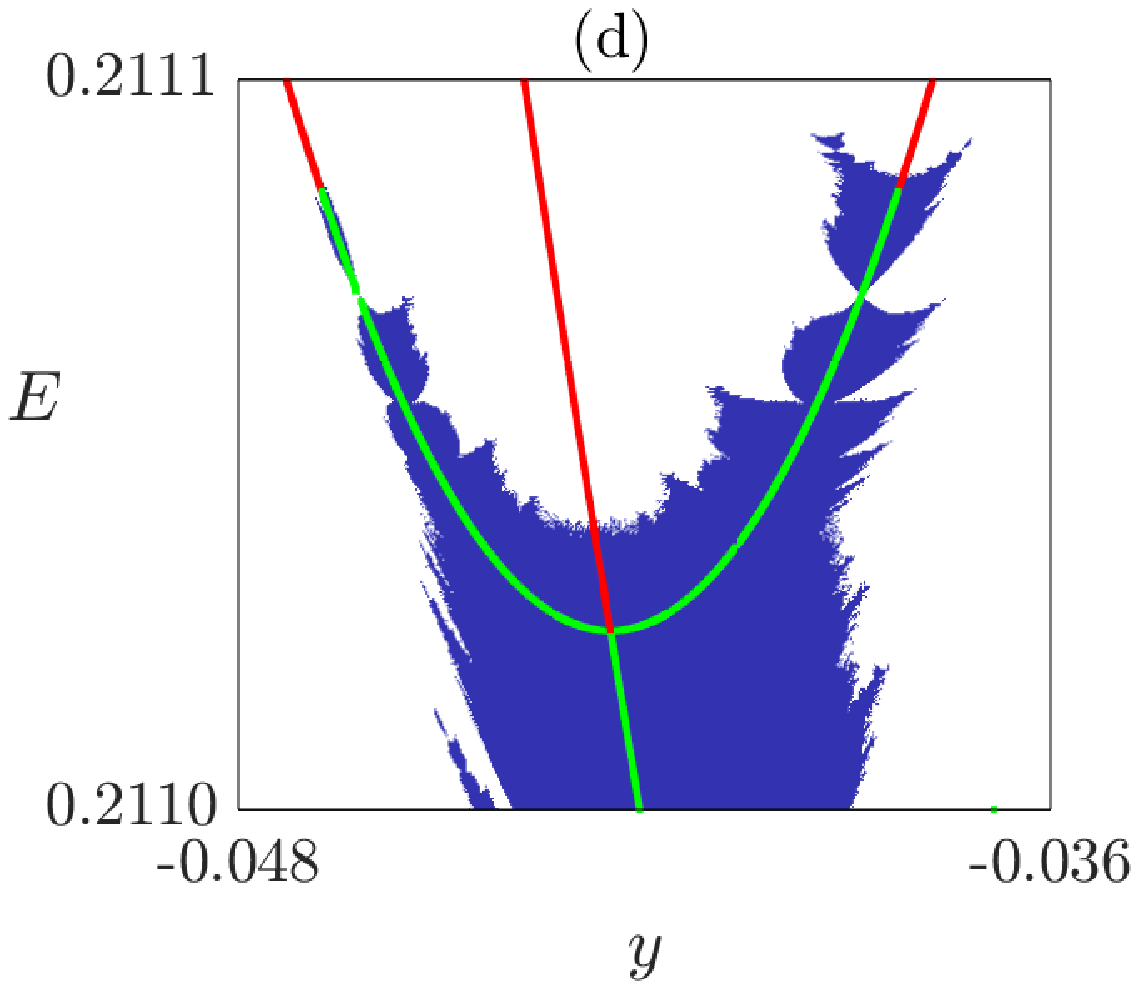}
	\caption{Branches of periodic orbits and KAM islands in the Hénon-Heiles system. The stable (unstable) periodic orbits are represented using green (red) lines. The KAM islands have been determined by computing the exit basins along the $y$-axis for different energies. Escaping initial conditions are colored in white, while KAM islands (non-escaping initial conditions) are represented in blue. Panel (a) shows the $m=2$ branches, while the next panels represent the subsequent period-doubling bifurcations. Note that each panel is a magnification of the area enclosed by dashed lines in the previous one.}
	\label{Fig4}
\end{figure}

The period-doubling bifurcations and their effects on the structure of KAM islands can be visualized by representing the branches of periodic orbits over an exit basin diagram in the $(y,E)$ plane. Since we are not interested here in the fractal structures of the exit basins, we have assigned white color to all escaping trajectories, while KAM islands are depicted in blue. The result is shown in Fig.~\ref{Fig4}, where green (red) lines denote stable (unstable) periodic orbits. In this figure, each panel is a magnification of the area enclosed by dashed lines in the previous one. Therefore, the $m=2$ branches are represented in Fig.~\ref{Fig4}(a), while the following panels show the subsequent period-doubling bifurcations. Regardless of the energy range, it can be observed that panels (b) and (d) exhibit the same qualitative features, while panel (c) is a mirror image of the other panels. As a matter if fact, this self-similar fractal structure repeats itself indefinitely within a finite energy range. Moreover, the bifurcations that occur in the branches of periodic orbits before they become unstable repeat in the same sequence at different scales. Therefore, each of these figures captures the fundamental aspects of the formation, evolution, and destruction of the main KAM island. We highlight that these structures are not representative of the Hénon-Heiles system only, but they are astonishingly similar in many different conservative systems (e.g, see Fig. 8 in \cite{Greene81} and Figs.~\ref{Fig9} and \ref{Fig10} of this manuscript).

\newpage
By detecting the loss of stability of periodic orbits, we have obtained numerically the energy values $E_n$ $(n=1,2,3...)$ where the first $7$ period-doubling bifurcations occur. The results are shown in the first three columns of Table~\ref{T1}. In this table, and throughout the whole manuscript, the uncertainty in the last significant digits of the parameters is indicated between parentheses. In the case of $E_n$, the uncertainty is given by half the difference between two consecutive energy values where we detect that the stability of the periodic orbit changes. 

Once we have obtained the parameter values where the period-doubling bifurcations occur, we can estimate the Feigenbaum constant, which is given by:
\begin{equation}
	\delta_H=\lim_{n\to\infty}\frac{E_{{n-1}}-E_{{n-2}}}{E_{{n}}-E_{{n-1}}},
\end{equation}
where the index $H$ indicates that the constant is calculated in a Hamiltonian system.

All estimates of $\delta_H$ are shown in the last column of Table~\ref{T1}, while the standard methods to calculate its uncertainty are explained in Appendix A. Our best approximation (using $E_5$, $E_6$, and $E_7$) is $\delta_H=8.72113(47)$, which agrees to a large extent with the result obtained by Greene \textit{et al.} in two-dimensional area-preserving maps \cite{Greene81} and by Mao \textit{et al.} in four-dimensional volume-preserving maps \cite{Mao85}. Therefore, we confirm that the value of the Feingenbaum constant is not only universal for area-preserving maps, but also for two-degree-of-freedom Hamiltonian systems. 

\begin{table}[h!]
	\begin{tabular}{c c c c} 
		\hline
		$n$ & $m$ & $E_n$ & $\delta_H$ \\ [0.5ex] 
		\hline\hline
		$1$  & $2$ & $0.1486504275(5)$ & - \\ 
		\hline
		$2$ & $4$ & $0.2062564235(5)$ & - \\
		\hline
		$3$ & $8$ & $0.2105406495(5)$ & $13.4460684(34)$ \\
		\hline
		$4$ & $16$ & $0.2110432870(1)$ & $8.523491(12)$ \\
		\hline
		$5$ & $32$ & $0.21110070066(4)$ & $8.754667(32)$ \\  
		\hline
		$6$ & $64$ & $0.21110728629(1)$ & $8.71802(87)$ \\
		\hline
		$7$ &$\quad 128 \quad$ & $0.211108041425(25)$ & $\quad 8.72113(47) \quad$ \\
		\hline		
	\end{tabular}
\caption{Values of the energy, $E_n$, where the first $7$ period-doubling bifurcations occur, together with estimations of the Feigenbaum constant $\delta_H$ using the former and the two previous values of $E_n$. The first two columns indicate the number of the period-doubling bifurcation and the multiplicity of the created periodic orbit, respectively. }
\label{T1}
\end{table}

The infinite sequence of period-doubling bifurcations occurs within a finite energy range. Therefore, exists an accumulation point that can be calculated as follows:
\begin{equation} 
	\begin{aligned}
		E_\infty & = E_{6}+\sum_{k=0}^{\infty}(E_{7+k}-E_{6+k})=E_6+\sum_{k=0}^{\infty}\frac{(E_7-E_6)}{\delta_H^k} \\
		& = E_6+\frac{\delta_H(E_7-E_6)}{\delta_H-1}=0.211108139226(35),
	\end{aligned}
\end{equation}
where we have used our best estimation for $\delta_H$. Using a more accurate value $\delta_H=8.721097200(1)$, we obtain $E_\infty=0.211108139227(30)$. Both estimations only differ in the last significant digit.

  \section{Islets of stability}\label{sec4}
\begin{figure}[h!]
  	\centering
  	\includegraphics[clip,height=9cm,trim=0cm 0cm 0cm 0cm]{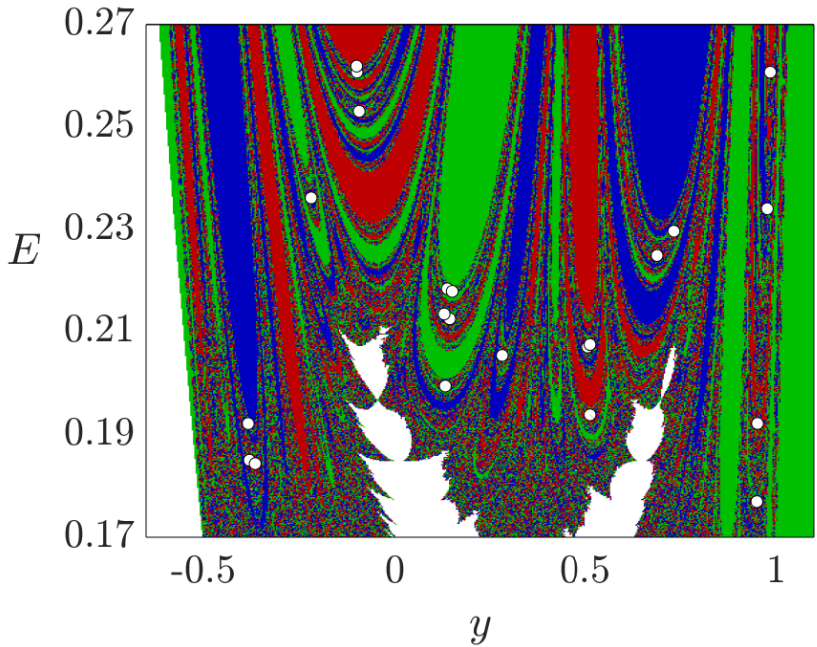}
  	\caption{Islets of stability (solid white dots) in an exit basin diagram for the Hénon-Heiles system. The color-code is as shown in the caption of Fig.~\ref{Fig2}. Note that the white region into the left part of the figure is a set of energetically forbidden initial conditions, not a KAM island.}
  	\label{Fig5}
  \end{figure}

  Although the only large KAM tori appear surrounding the main family of periodic orbits, unrelated and occasionally stable branches generate islets of stability. Since all periodic orbits cross at least once the $(0,y,\dot{x}(y,E),0)$ Poincaré section, we can ensure that islets will appear on the $(y,E)$ exit basin diagram. Furthermore, as periodic orbits make up the boundary of the exit basins, the search for islets can be constrained. Following these facts, we have found $24$ of them by performing a detailed grid search out in the boundary of the exit basins. Of course, by delving further into the structure of the boundary, one may discover an arbitrarily large number of islets. Due to their reduced area in the $(y,E)$ plane, we indicate their position by using solid white dots in Fig.~\ref{Fig5}. In this figure, the $m=2$ branches of the main KAM island can be clearly observed at the bottom of the plot (note that, colors aside, Fig.~\ref{Fig4}(a) is a magnification of Fig.~\ref{Fig5} in the vicinity of the main KAM island).     
  
\newpage

\begin{figure}[h!]
	\centering
	\includegraphics[clip,height=6cm,trim=0cm 0cm 0cm 0cm]{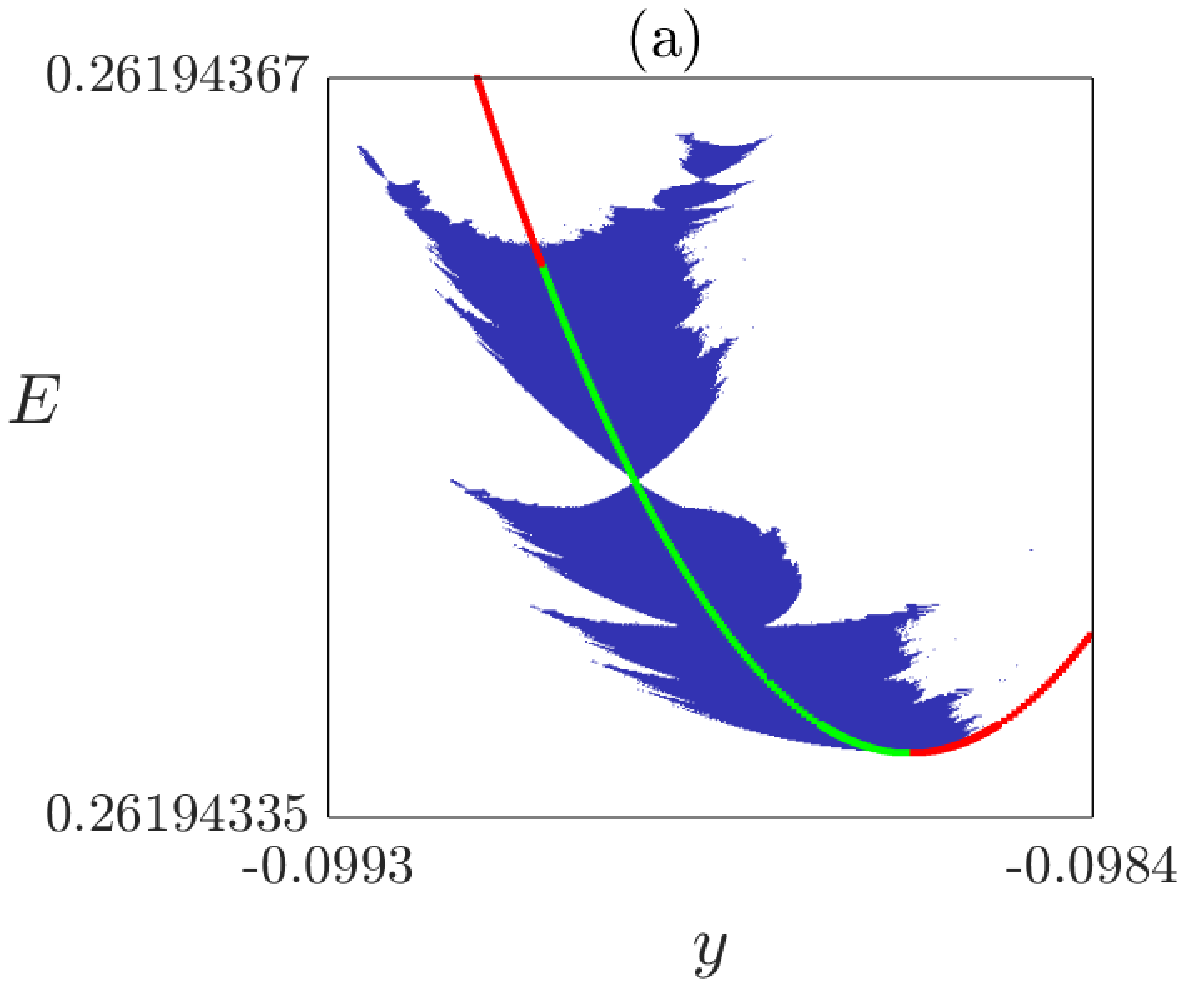}
	\includegraphics[clip,height=6cm,trim=0cm 0cm 0cm 0cm]{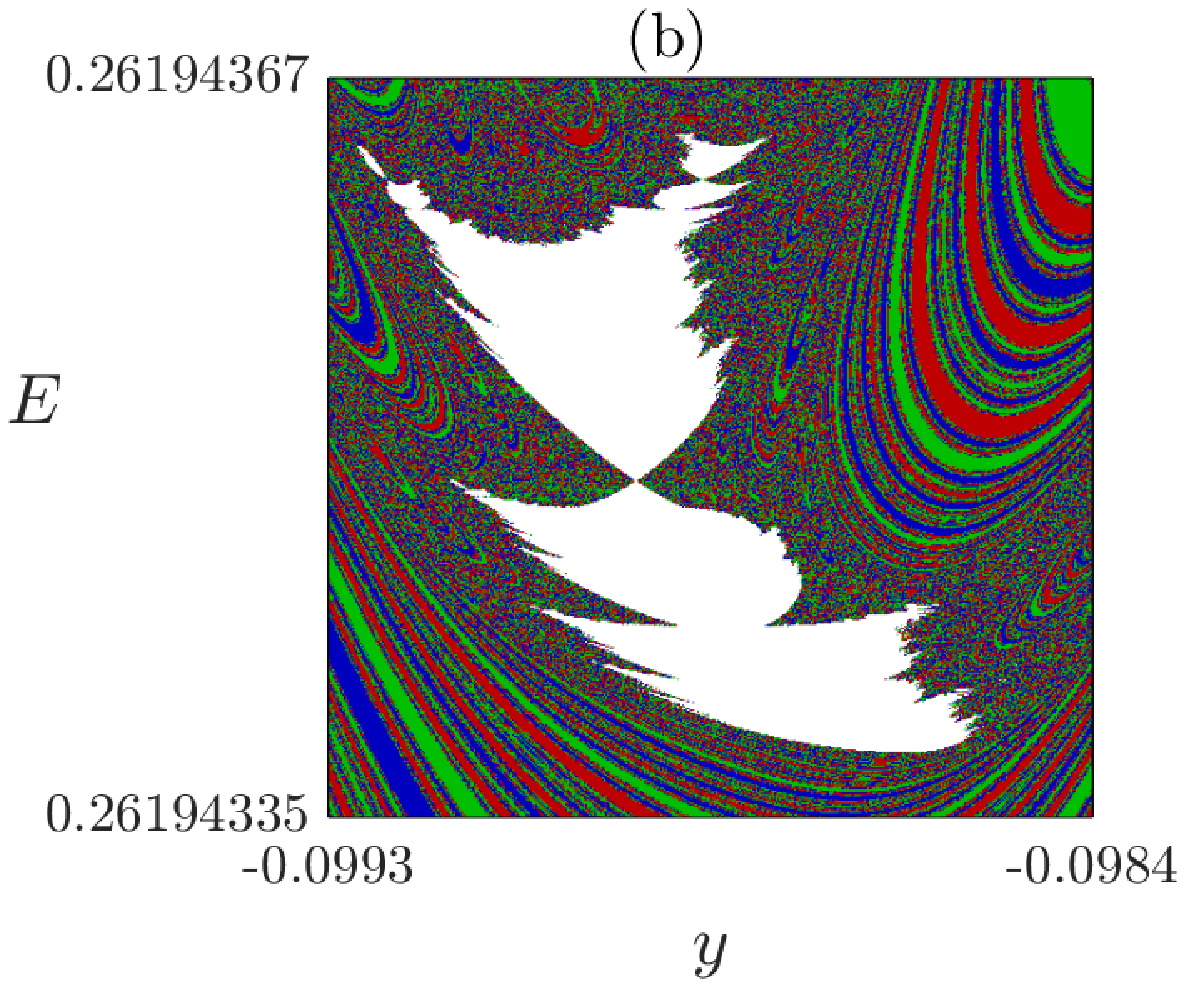}
	\includegraphics[clip,height=6cm,trim=0cm 0cm 0cm 0cm]{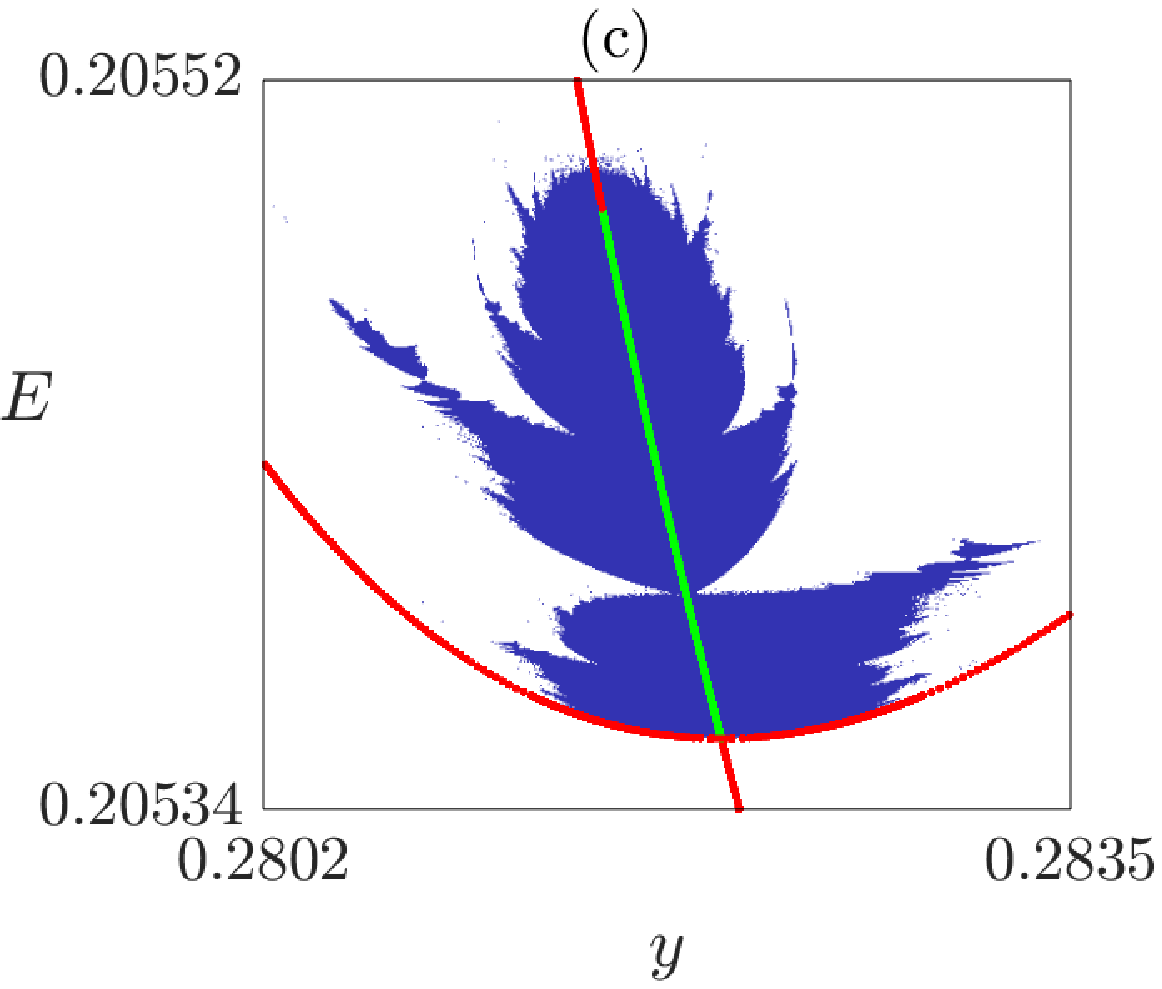}
	\includegraphics[clip,height=6cm,trim=0cm 0cm 0cm 0cm]{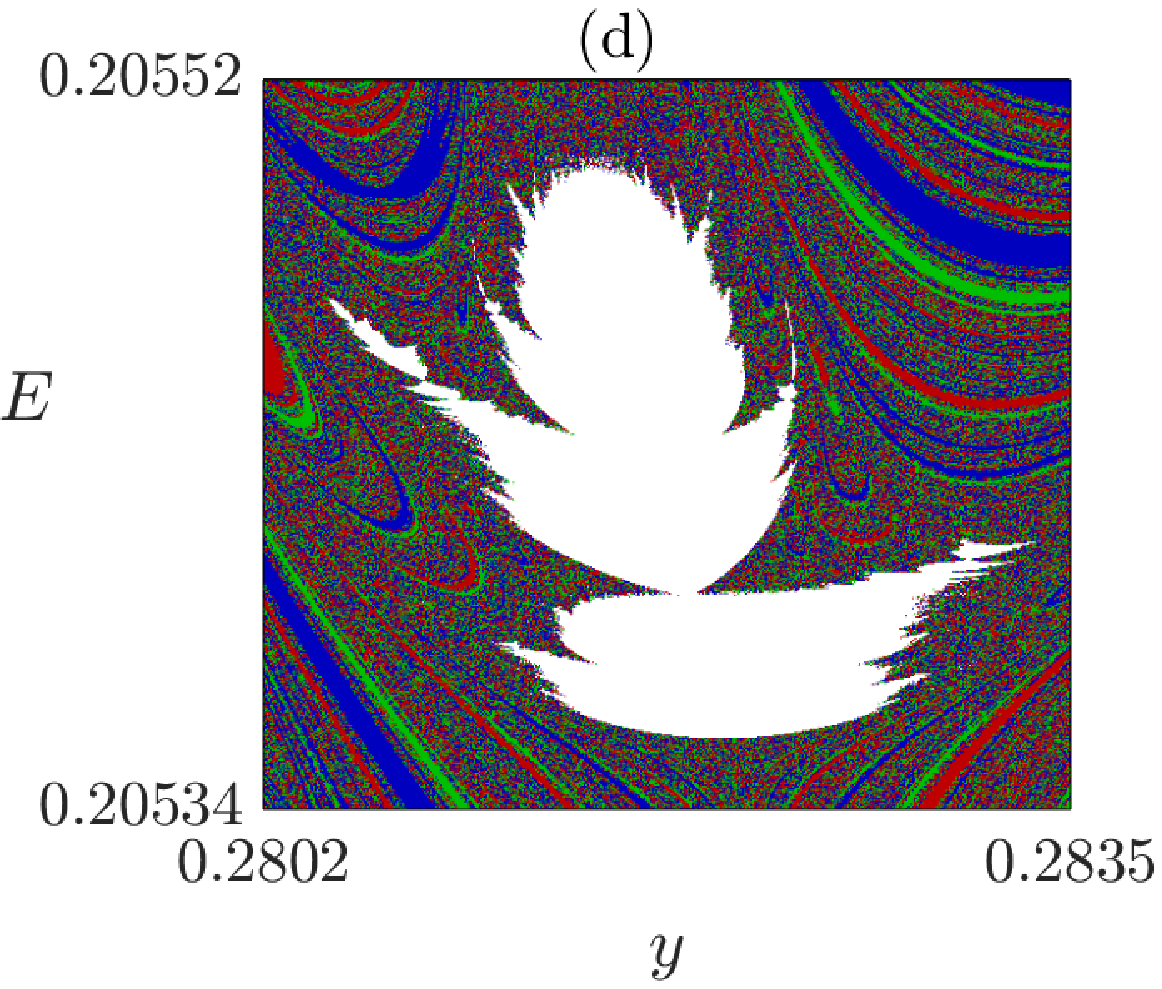}
	\includegraphics[clip,height=6cm,trim=0cm 0cm 0cm 0cm]{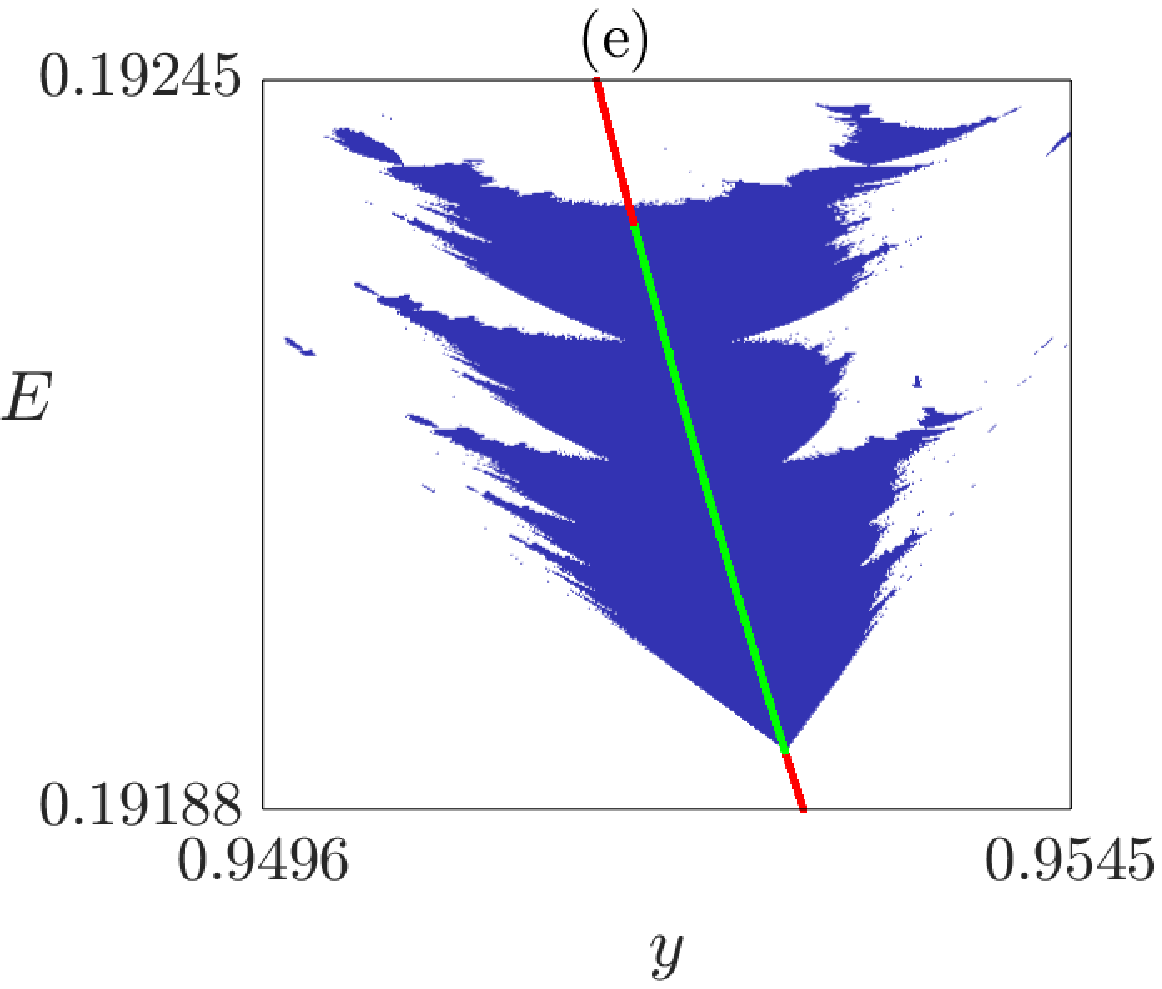}
	\includegraphics[clip,height=6cm,trim=0cm 0cm 0cm 0cm]{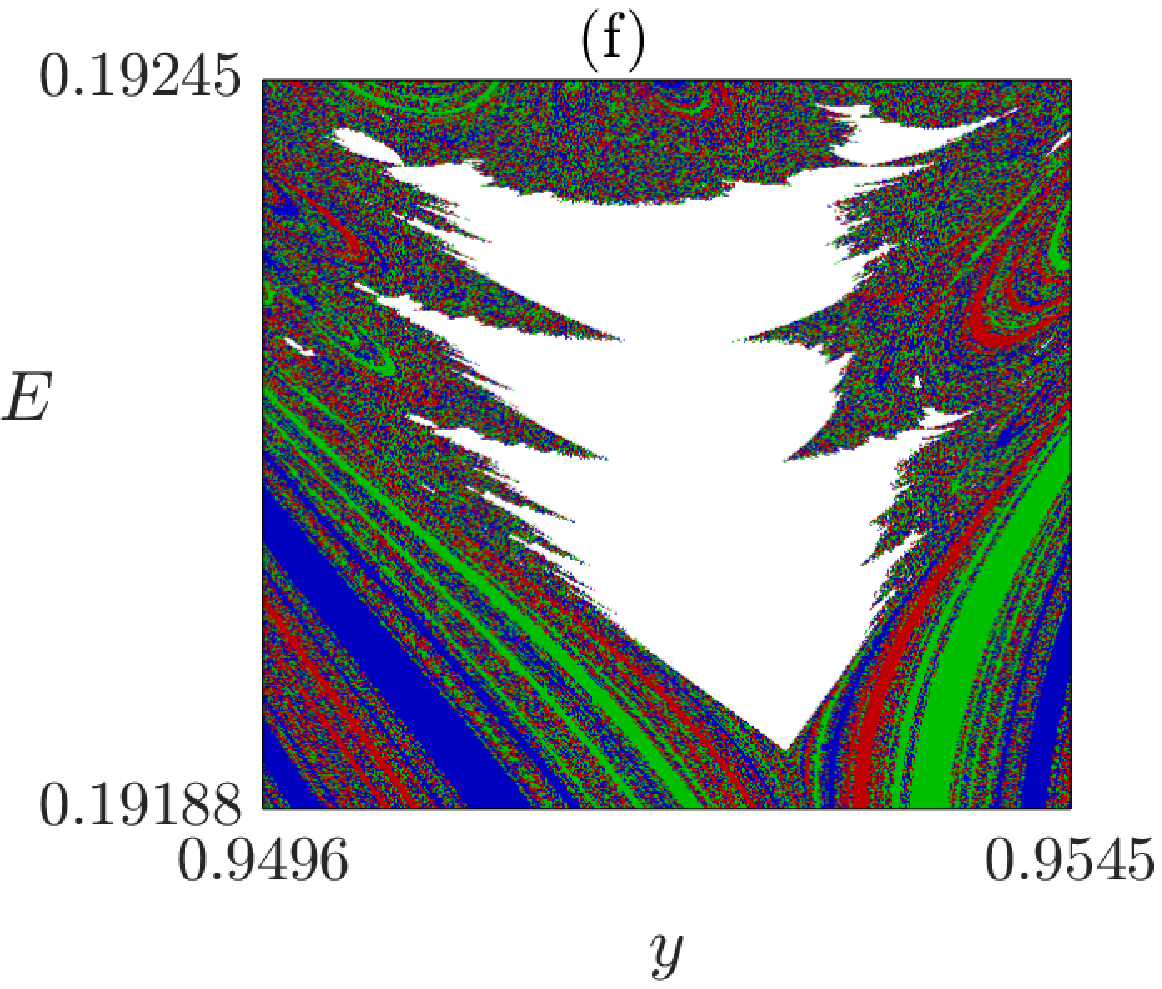}
	\caption{Representative examples of the different types of islets. The pairs of panels (a-b), (c-d), and (e-f) represent islets of types I, II, and III, respectively. These pairs of panels contain similar information, but from different perspectives. Panels (a,c,e) display the bifurcations and the emergence of islets surrounding stable periodic orbits. Panels (b,d,f) represent the islets in contrast to the fractal basin boundary. In panels (a,c,e) the color-code is as in Fig.~\ref{Fig4}, while in panels (b,d,f) is as in Fig.~\ref{Fig2}.} 
	\label{Fig6}
\end{figure}

We have studied individually each detected islet and, based on the bifurcations of periodic orbits involved in their formation, we have classified them into three different types. For a better understanding of their origin, we can observe that they appear near the edge of the parabolic shapes arising in the basin boundary (see Fig.~\ref{Fig5}). These parabolic shapes correspond to an infinite set of bifurcations, usually characterized by the birth of two unstable branches which correspond to a single unstable periodic orbit that crosses the Poincaré section twice. Nonetheless, in some cases a pair of stable-unstable periodic orbits is created in a saddle-node bifurcation. The stable branch is the responsible of the formation of a type I islet (see Fig.~\ref{Fig6}(a-b)). The remaining two types of islets always appear in branches of periodic orbits created in a saddle-node bifurcation. Therefore, islets of types II and III are always preceded by a type I islet. 

The stable periodic orbit that generates a type I islet eventually loses its stability after undergoing some standard bifurcation (typically pitchfork). For slightly higher energy values, the periodic orbit can become stable again, creating a type II islet (see Fig.~\ref{Fig6}(c-d)). Hence, if a type II islet exists, it always appears in the same branch where a type I islet existed (i.e., in the stable branch created in the saddle-node bifurcation). However, we emphasize that not all type I islets are followed by a type II islet, but they can also be alone.
  
 Type III islets can appear in both branches that are created in the saddle-node bifurcation. They arise from bifurcations where an unstable periodic orbit becomes stable (see Fig.~\ref{Fig6}(e-d)). While type II islets exhibit a smooth shape near the bifurcation point, type III islets are characterized by a sharp edge. Unlike the previous types, we have not observed the emergence of new unstable periodic orbits in the bifurcation leading to type III islets.
  
  For the sake of reproducibility, in Table~\ref{T2} (see Appendix B) we list the range of coordinates in the $(y,E)$ plane where the $24$ islets that we have detected can be found. We also indicate their type and the multiplicity of the generating periodic orbit. Except for the $24$th islet, we have detected and listed the islets that occupy a bigger area in the $(y,E)$ plane (in the case of the $24$th islet we have used higher resolution in the exit basin diagram with the aim of finding the energy value which generates the last islet). As can be seen in Table~\ref{T2}, the periodic orbits have a relatively low multiplicity. This fact suggests that periodic orbits with high multiplicity generate smaller islets.  
  
\begin{figure}[h!]
  	\centering
  	\includegraphics[clip,height=4.7cm,trim=0cm 0cm 0cm 0cm]{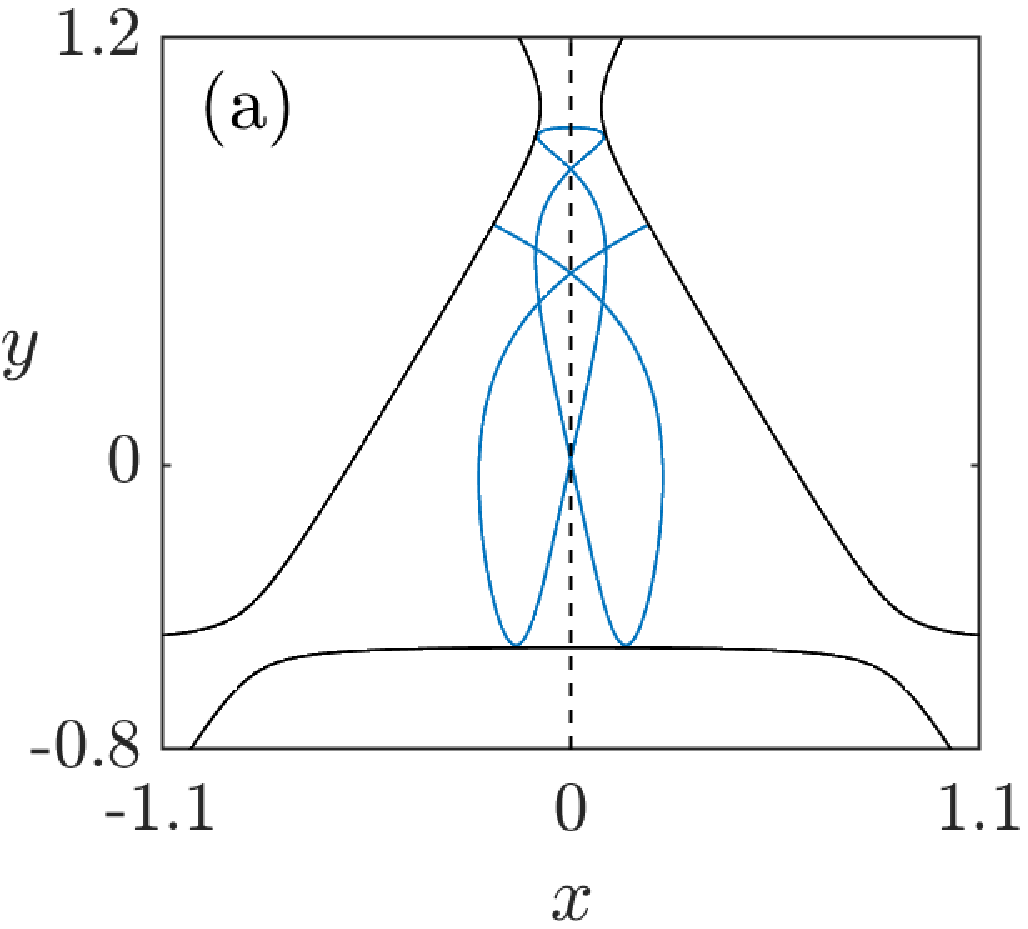}
  	\includegraphics[clip,height=4.7cm,trim=0cm 0cm 0cm 0cm]{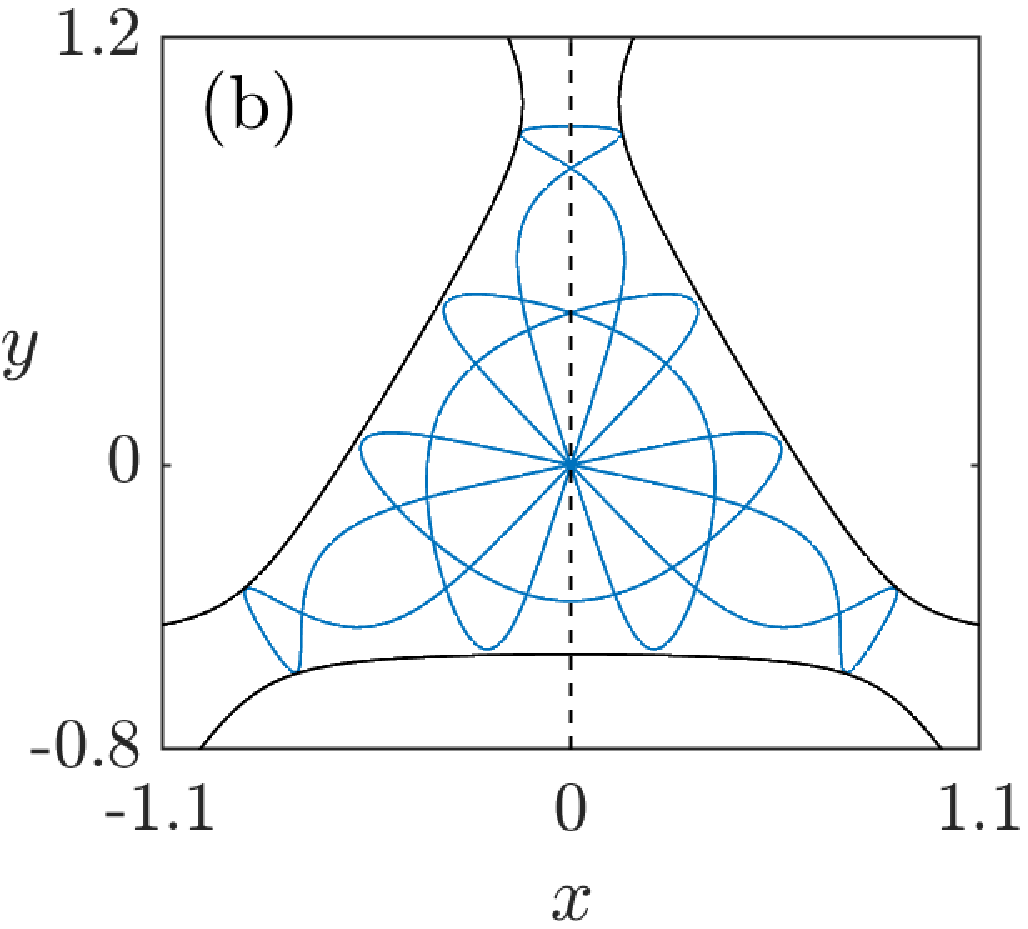}
  	\includegraphics[clip,height=4.7cm,trim=0cm 0cm 0cm 0cm]{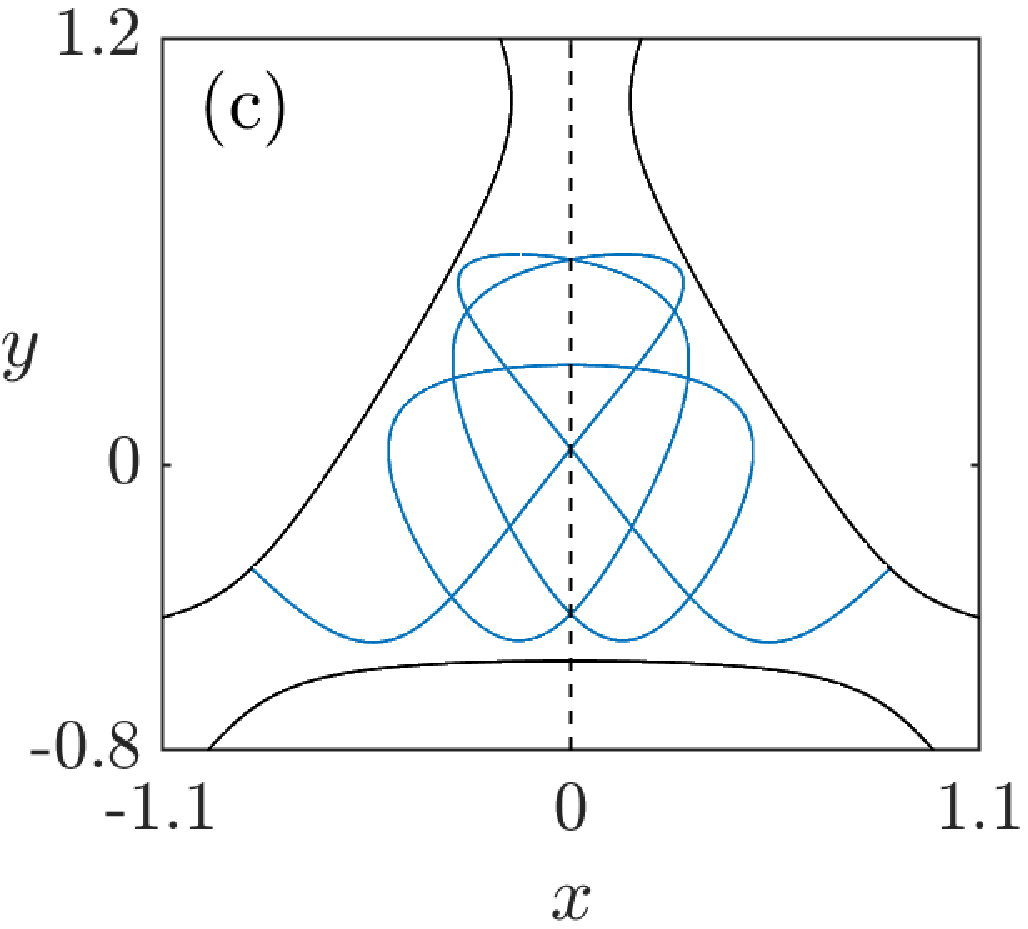}
  	\includegraphics[clip,height=4.7cm,trim=0cm 0cm 0cm 0cm]{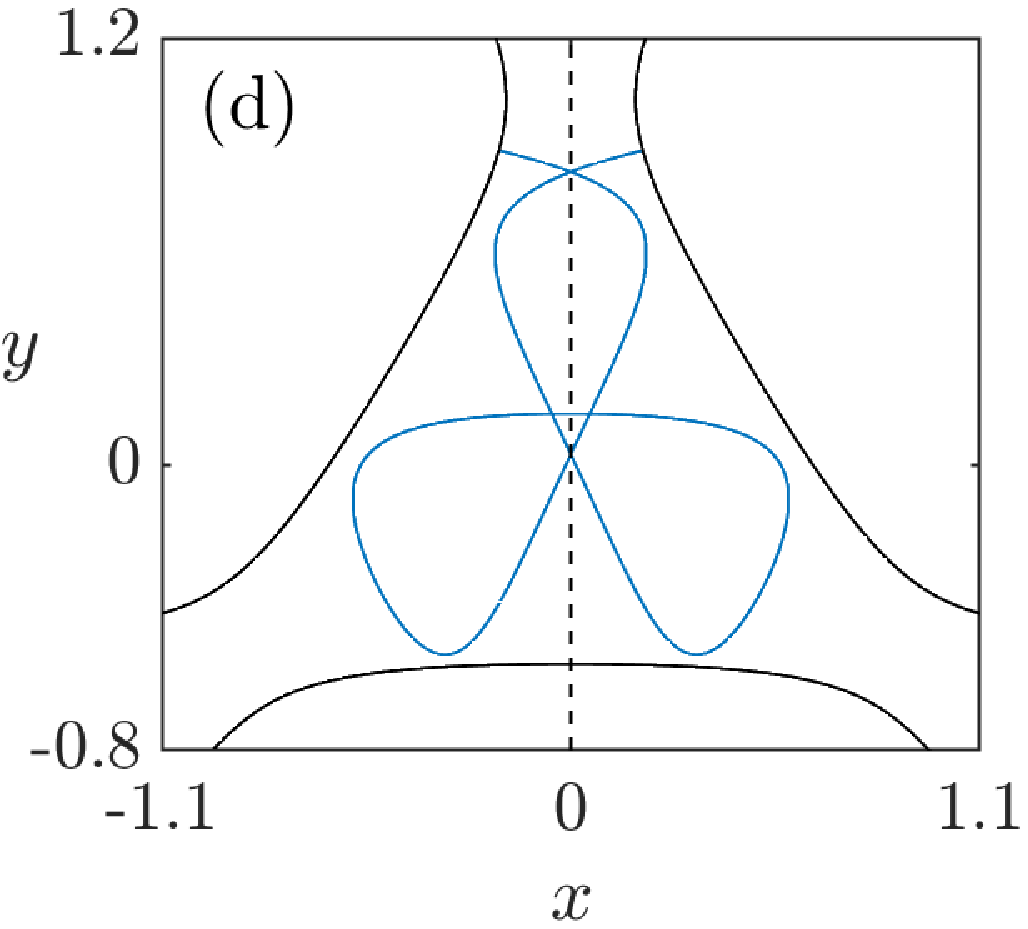}
  	\includegraphics[clip,height=4.7cm,trim=0cm 0cm 0cm 0cm]{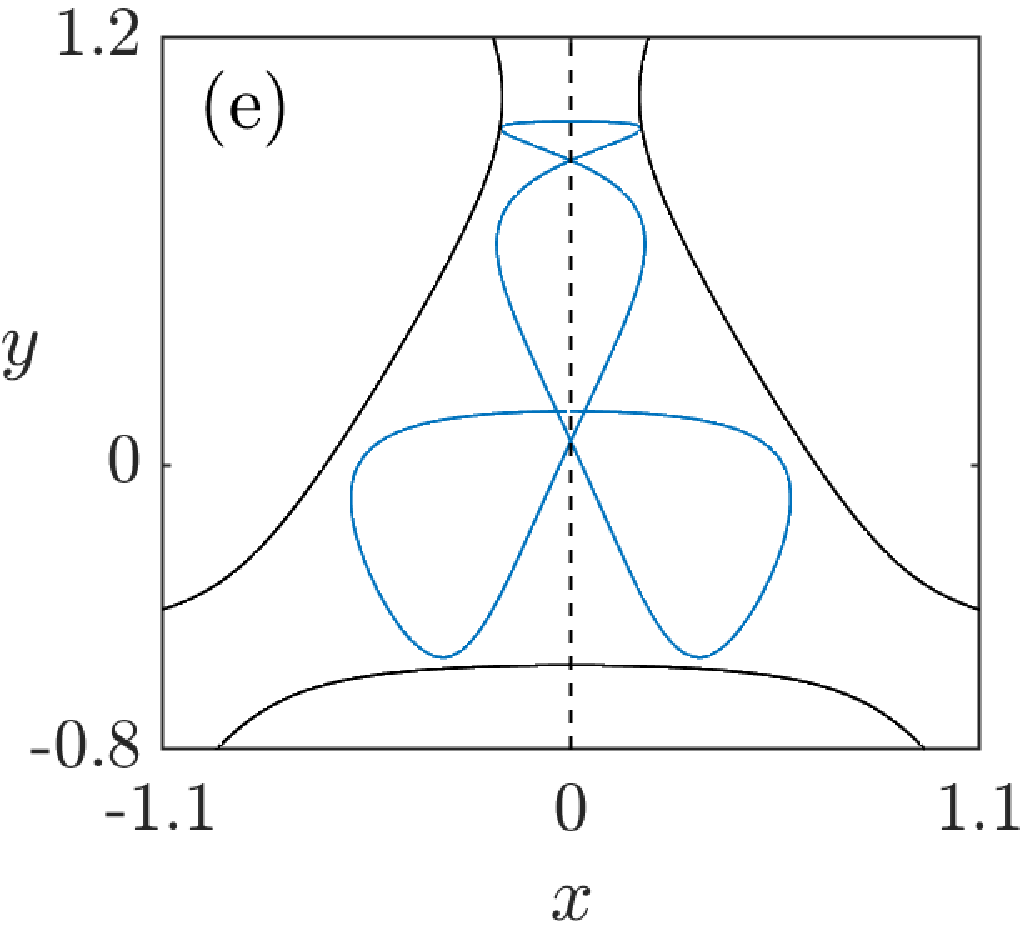}
  	\includegraphics[clip,height=4.7cm,trim=0cm 0cm 0cm 0cm]{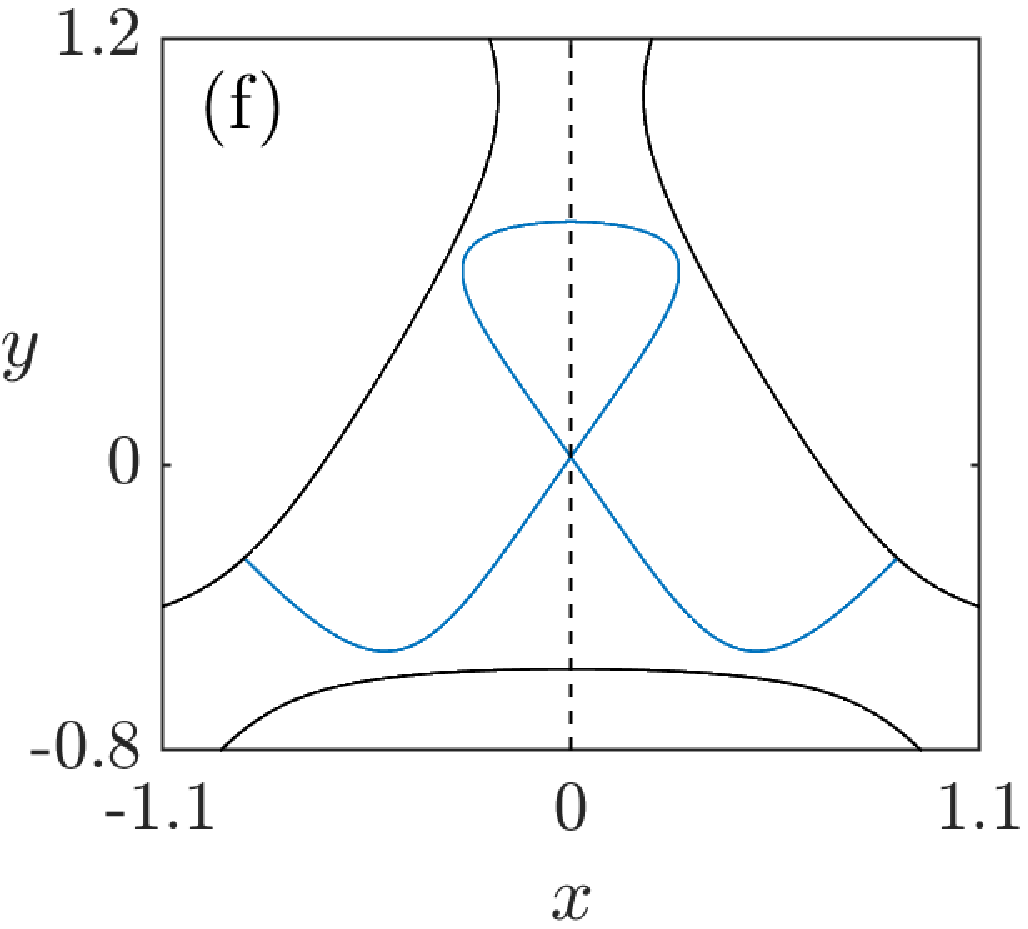}
  	\includegraphics[clip,height=4.7cm,trim=0cm 0cm 0cm 0cm]{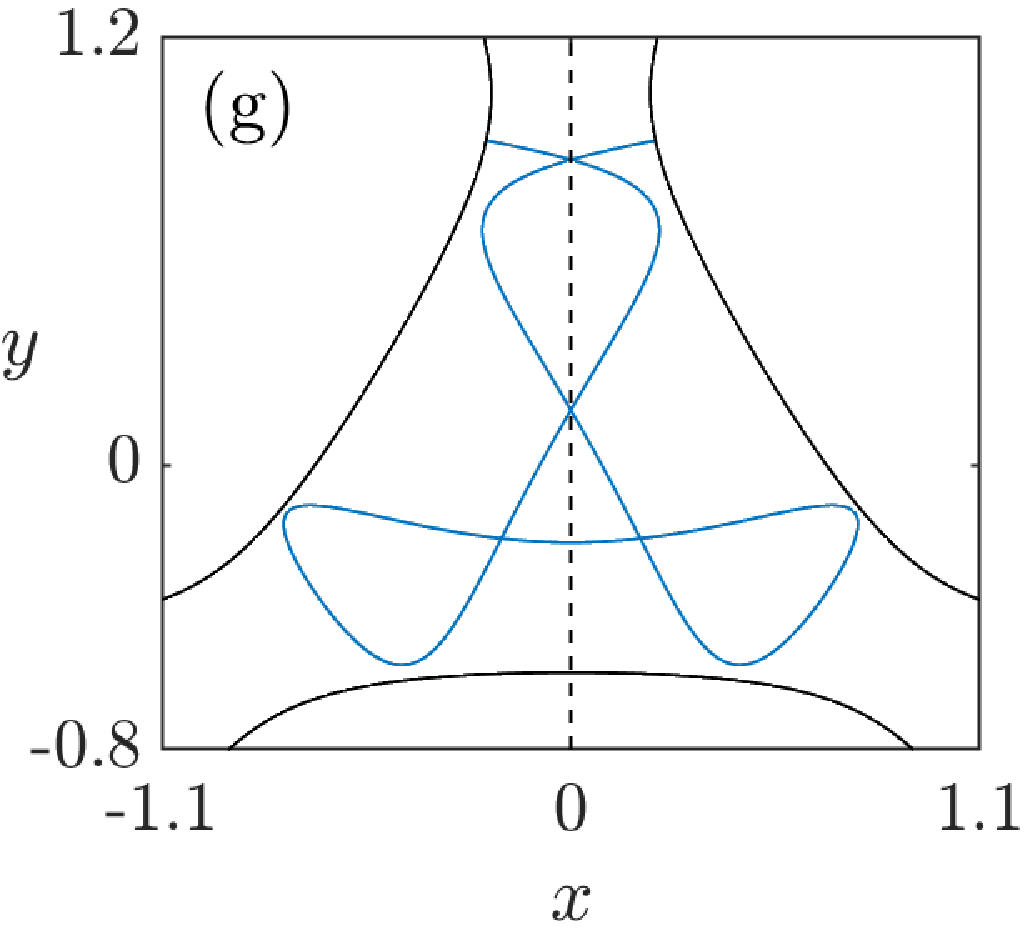}
  	\includegraphics[clip,height=4.7cm,trim=0cm 0cm 0cm 0cm]{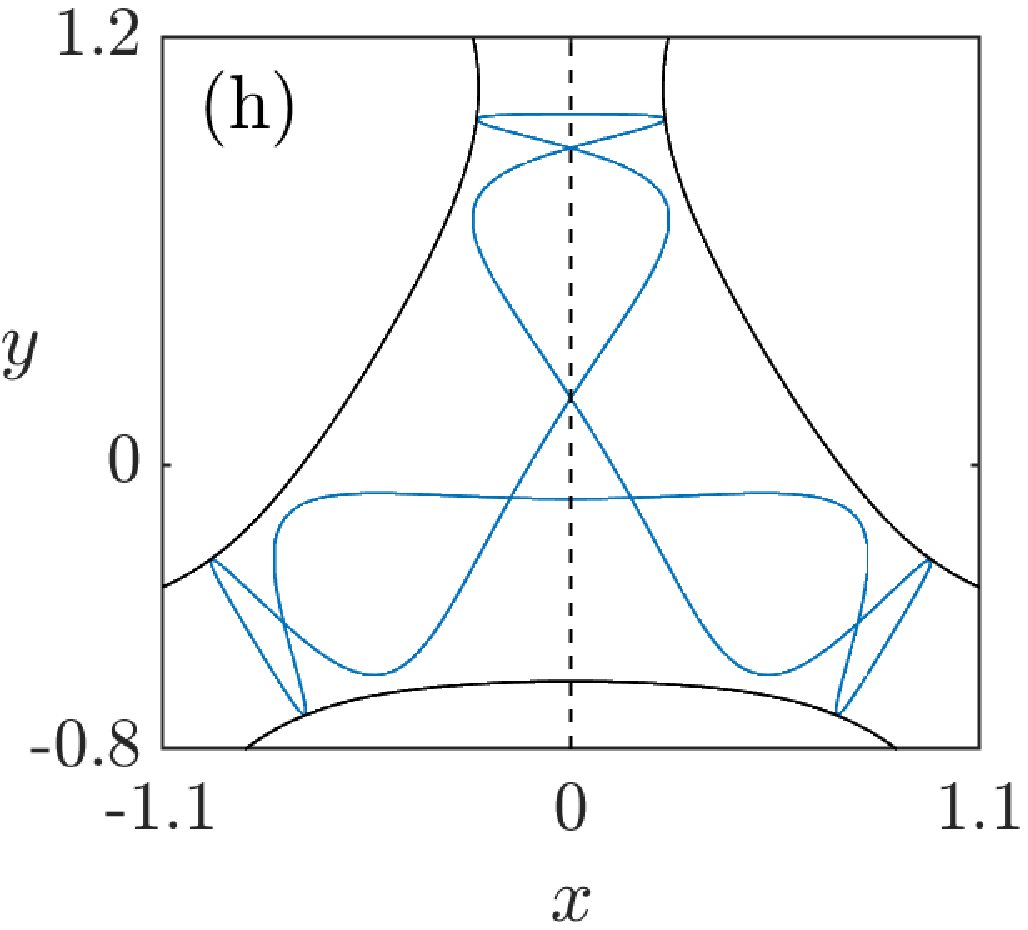}
  	\includegraphics[clip,height=4.7cm,trim=0cm 0cm 0cm 0cm]{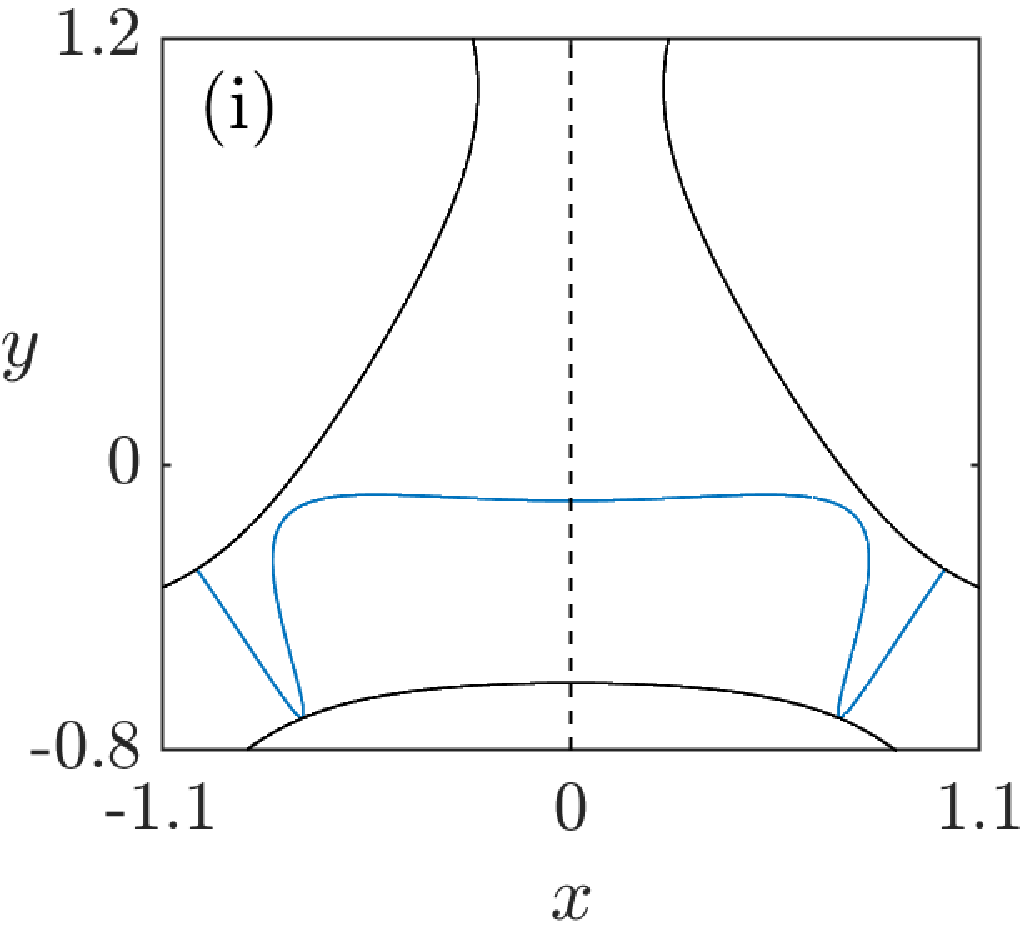}
  	\caption{A gallery of stable periodic orbits for the Hénon-Heiles system. Each of these orbits generates one of the $24$ islets that we have detected, classified, and listed in Table~\ref{T1} (Appendix A). In particular, the number of the corresponding islet is (a) $1$, (b) $5$ and $6$, (c) $9$, (d) $12$, (e) $14$, (f) $16$, (g) $19$, (h) $22$ and $23$, and (i) $24$. }
  	\label{Fig7}
  \end{figure}

  For illustrative purposes, in Fig.~\ref{Fig7} we represent in the $(x,y)$ plane some stable periodic orbits that generate islets. Note that a single periodic orbit can cross the $(0,y,\dot{x}(y,E),0)$ Poincaré section twice (e.g., the periodic orbits represented in panels (a) and (h) in Fig.~\ref{Fig7}). In these cases, two islets of the same type appear in the $(y,E)$ plane. 
     
\begin{figure}[h!]
  	\centering
  	\includegraphics[clip,height=6cm,trim=0cm 0cm 0cm 0cm]{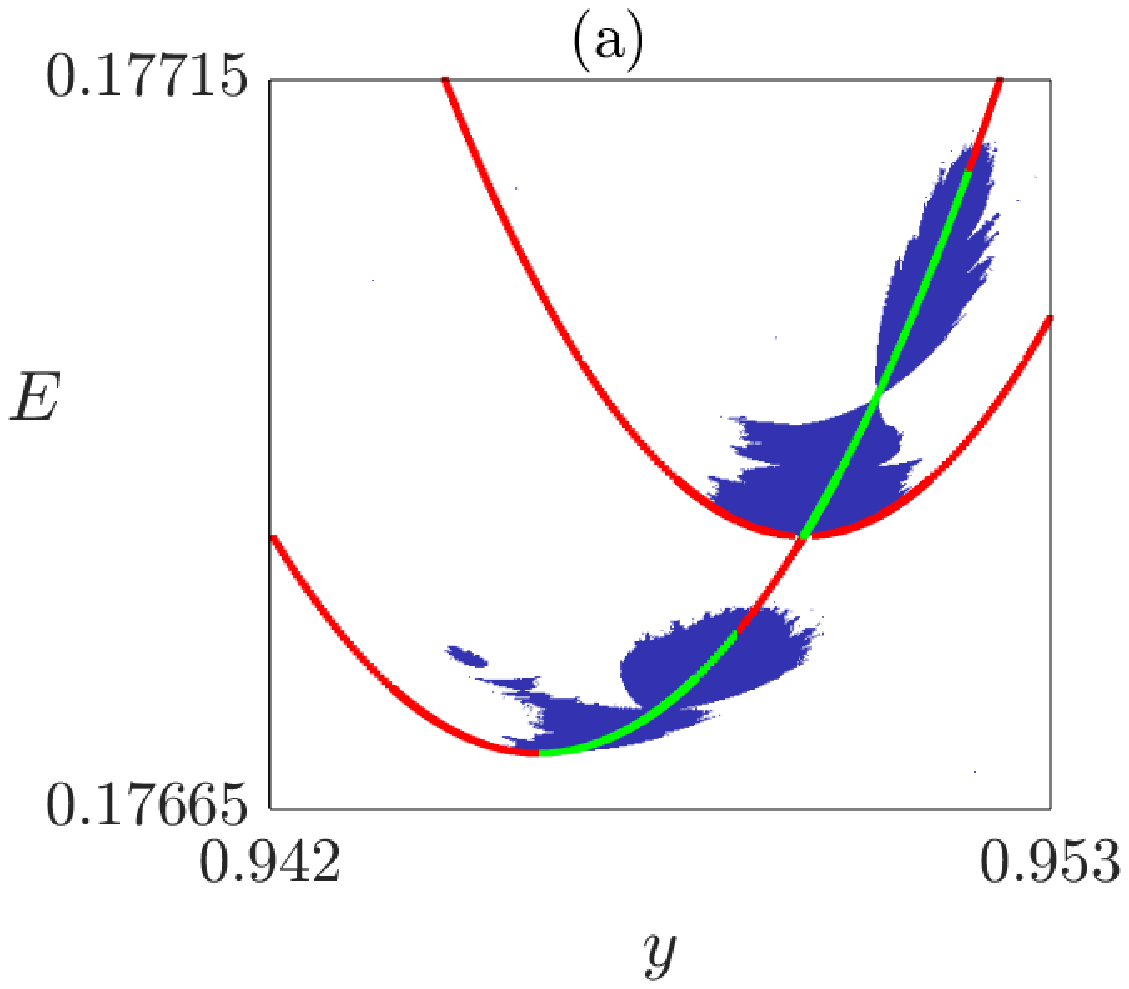}
  	\includegraphics[clip,height=6cm,trim=0cm 0cm 0cm 0cm]{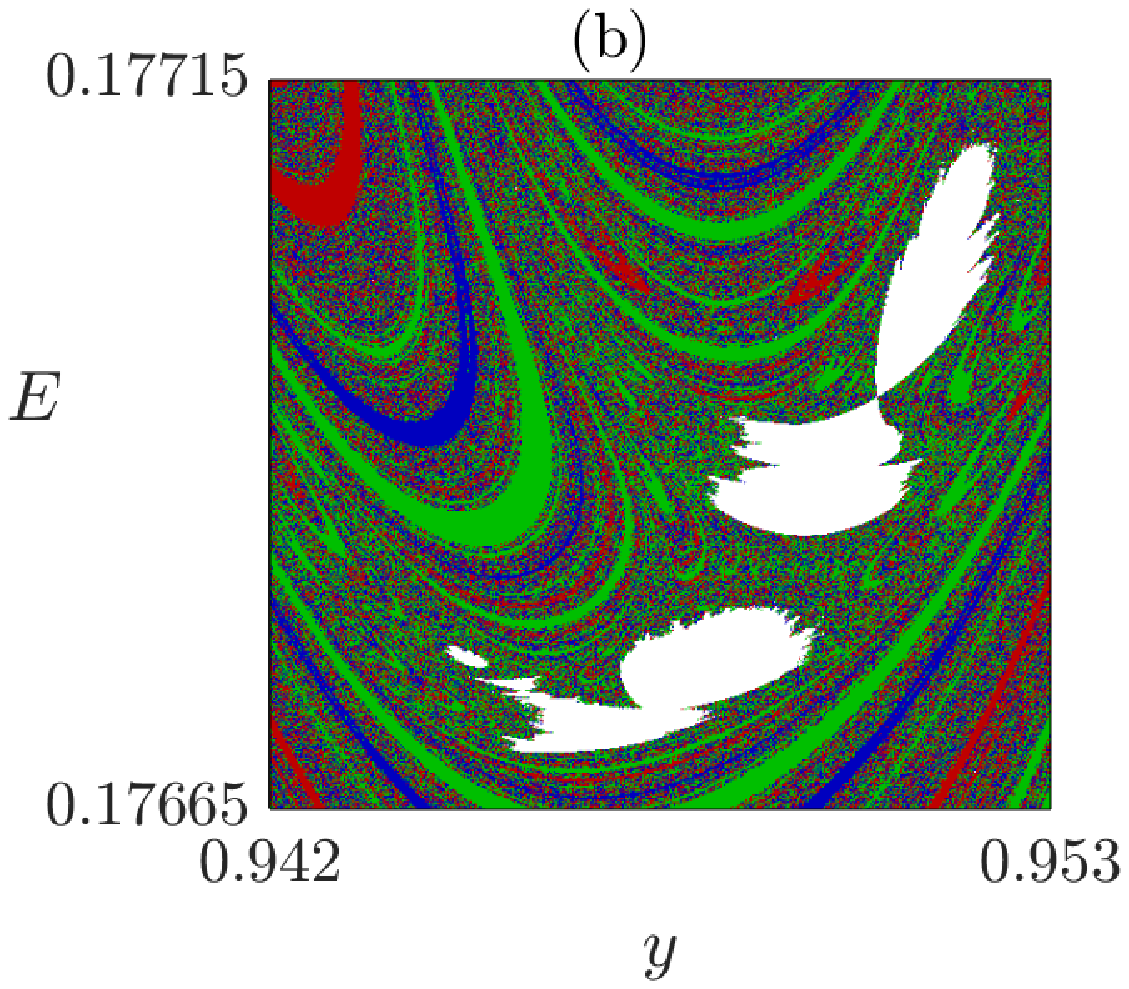}\\\vspace{0.4cm}
  	\includegraphics[clip,height=6cm,trim=0cm 0cm 0cm 0cm]{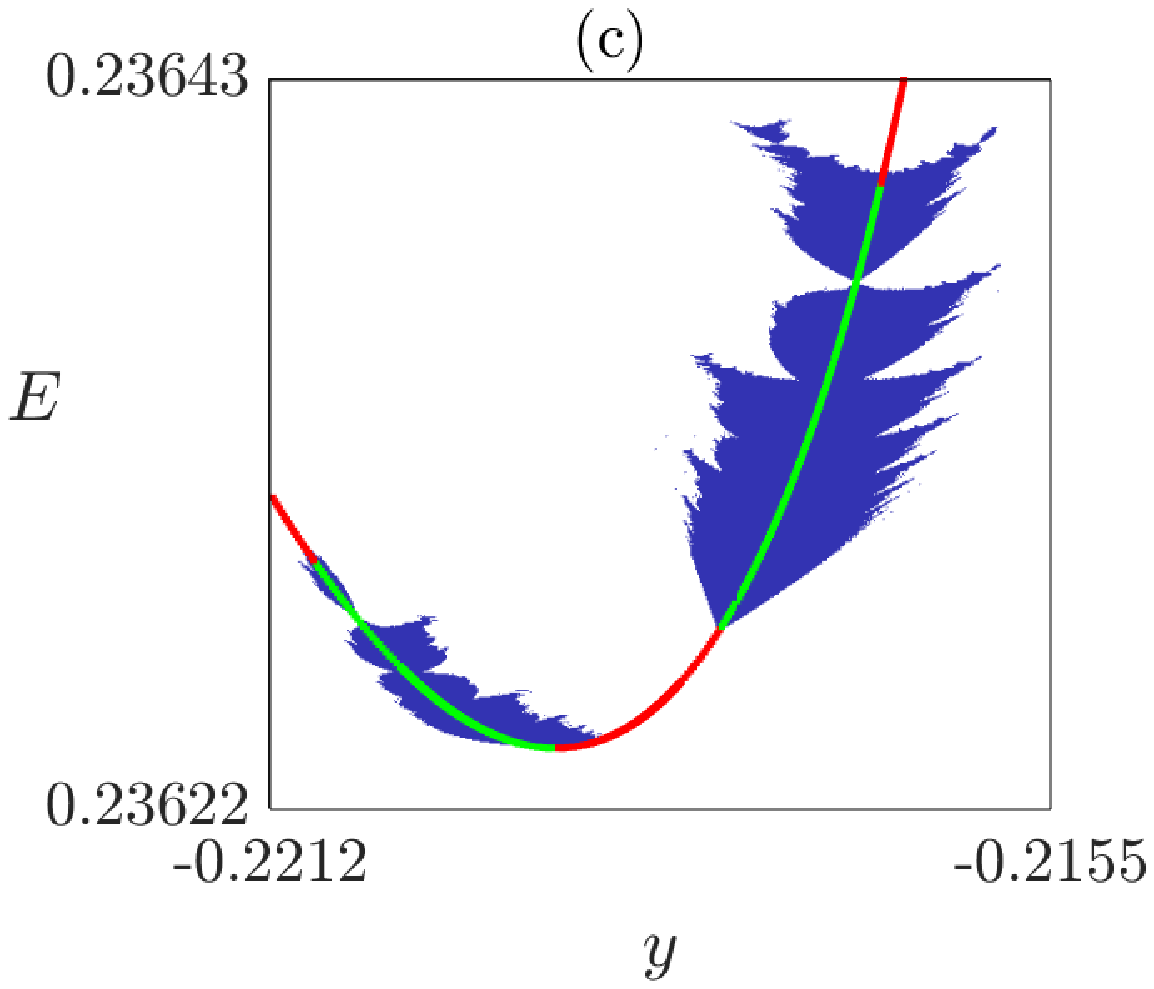}
  	\includegraphics[clip,height=6cm,trim=0cm 0cm 0cm 0cm]{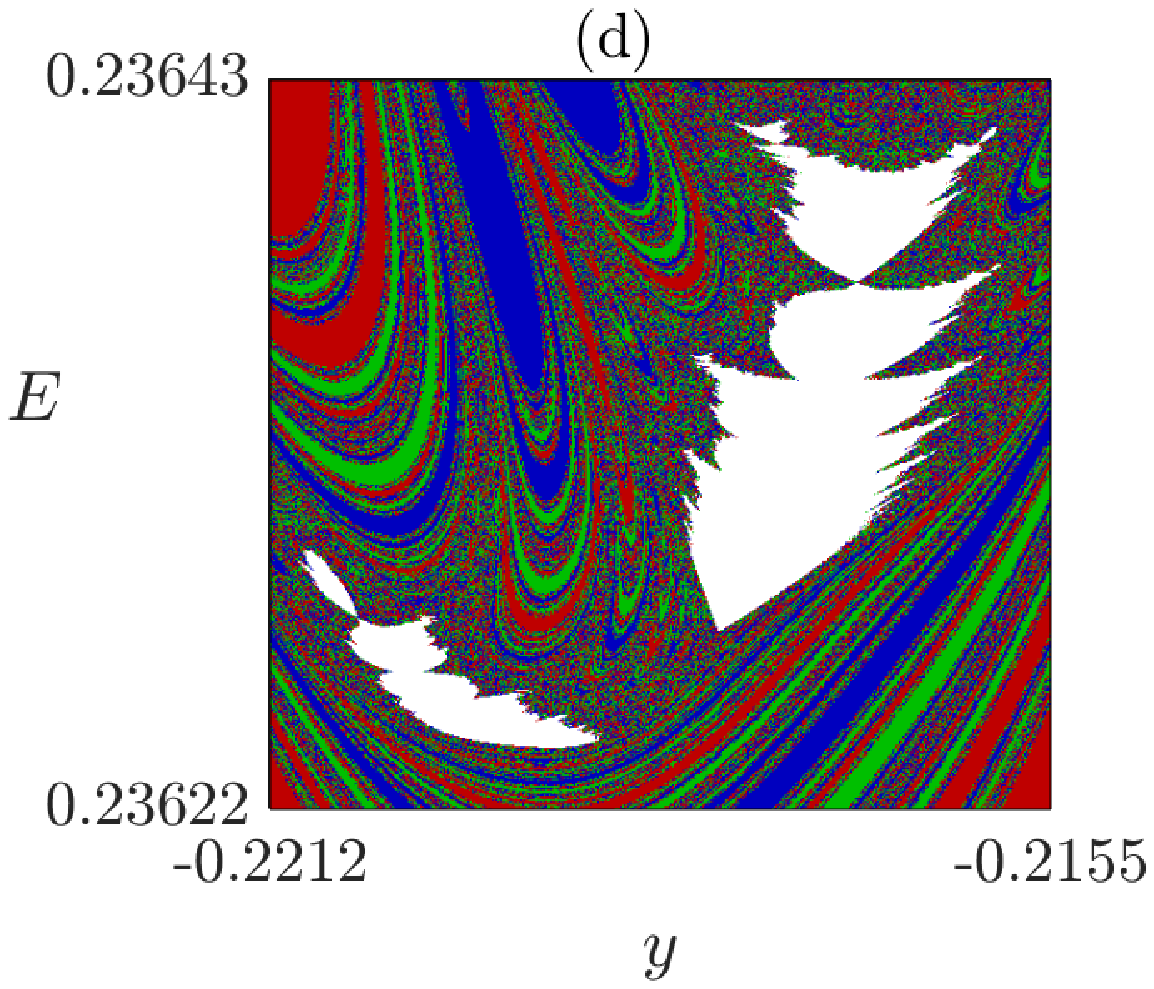}
  	\caption{Two examples where islets of different types appear within a reduced energy range. The pairs of panels (a-b) and (c-d) contain similar information, but from different perspectives. Panels (a,c) display the bifurcations and the emergence of islets surrounding stable periodic orbits. Panels (b,d) represent the islets in contrast to the fractal basin boundary. In panels (a,c) the color-code is as in Fig.~\ref{Fig4}, while in panels (b,d) is as in Fig.~\ref{Fig2}.} 
  	\label{Fig8}
  \end{figure}

  Occasionally, islets of types II and III can be observed in the same plot as type I islets, since they appear for close energy values. Two examples of this phenomenon are displayed in Fig.~\ref{Fig8}. In panels (a-b), we can see a type II islet forming in the same branch where a type I islet previously existed at lower energy levels. In panels (c-d), we see how a type III islet appears after the unstable branch created in a saddle-node bifurcation becomes occasionally stable. In this case, during a short energy range, islets of types I and III coexist.  
  
 \newpage
We aim to conclude our findings on the Hénon-Heiles system by discussing an aspect that attracted the attention of some researchers: the energy value $E_k$ for which the KAM tori disappear. Regarding this matter, various energy values have been put forward in the literature. The initial approximation to this limit value was $E_k\approx0.2113$ \cite{Barrio08}, which is a rough approximation of the accumulation point. Another suggested value was $E_k\approx0.2309$ \cite{Nieto20}, which probably arose as a result of detecting the islet number $17$ (see Table~\ref{T2}). Finally, a recent paper found an islet for $E_k\approx0.2534$ (islet number $21$ in Table~\ref{T2}). In our numerical simulations, the last detected islet is destroyed for $E_k\approx0.26194367$ (islet number $24$ in Table~\ref{T2}).

From the previous information, it is clear that the value of $E_k$ is gradually increased due to higher precision in the numerical simulations. This is not surprising, since the range of energies where islets appear is reduced as the energy of the system is increased. However, bifurcations do not occur for arbitrarily high values of the energy. After searching into the structure of the boundary of the exit basins, we have found that the last bifurcation occurs for $E=0.262158902577(1)$. We have not found a stable periodic orbit nor an islet in the neighborhood of the last bifurcation, but its existence cannot be definitively dismissed. Therefore, we cannot provide an exact value for $E_k$, but we conjecture that its value is not significantly above the energy where the last bifurcation occurs. 

\section{Islets of stability in different systems}\label{sec5}

The same types of islets that we have found in the Hénon-Heiles system appear in generic two-degree-of-freedom Hamiltonian systems and area-preserving maps. To illustrate this generality, in this section we provide numerical evidence of the existence of islets in the Barbanis system \cite{Barbanis} and in the standard map (also known as Chirikov-Taylor map) \cite{Chirikov}.

The Barbanis system is a two-degree-of-freedom Hamiltonian system given by:
\begin{equation}
	{\cal{H}}=\frac{1}{2}(\dot{x}^2+{\dot{y}}^2)+\frac{1}{2}(x^2+y^2)-xy^2.
\end{equation}

\begin{figure}[h!]
	\centering
	\includegraphics[clip,height=9cm,trim=0cm 0cm 0cm 0cm]{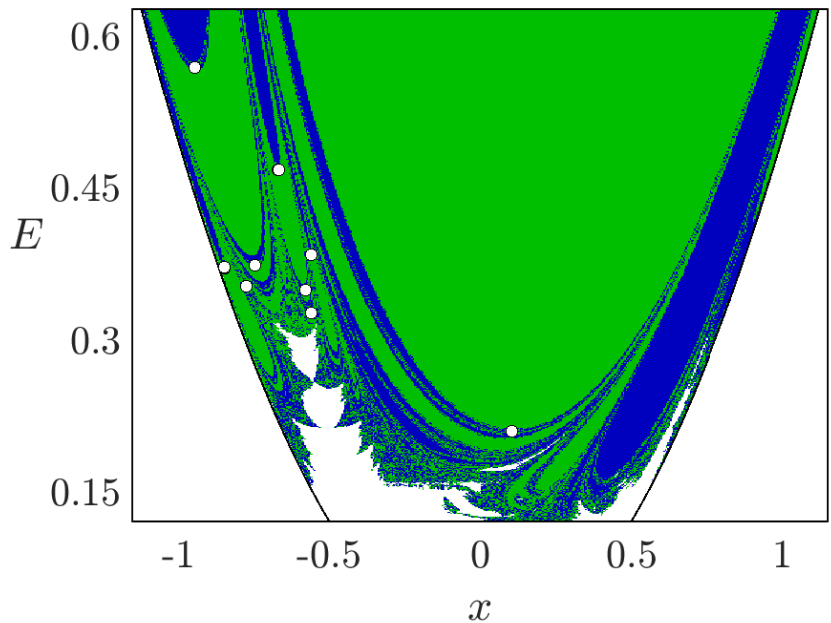}
	\caption{Islets of stability (solid white dots) in an exit basin diagram for the Barbanis system. The colors green and blue refer to initial conditions leading to the two exits of the potential: Exit $1$ ($y\to\infty$) and Exit $2$ ($y\to-\infty$). White regions inside the potential correspond to KAM islands. }
	\label{Fig9}
\end{figure}

Besides being time-reversible, the system is symmetric about the $x$-axis. Therefore, using similar arguments to those exposed in the Hénon-Heiles system, the condition for a periodic orbit to exist in the Barbanis system is $y(x_0,0,0,\dot{y}_0;T/2)=\dot{x}(x_0,0,0,\dot{y}_0;T/2)=0$. Thus, for detecting islets we have chosen the $(x,0,0,\dot{y})$ Poincaré section and we have computed an exit basin diagram in the $(x,E)$ plane. The result is shown in Fig.~\ref{Fig9}, where the position of $12$ islets is represented with white dots. Here we only see two colors in the exit basin diagram since the system exhibits two exits. The coordinate range where the islets can be found in this system is shown in Table~\ref{T3} (see Appendix B).

\newpage

On the other hand, the standard map is an area-preserving map defined by the following formula:
\begin{equation} 
	\begin{aligned}
		\theta_{n+1} & = \theta_n+J_{n+1} \quad \mbox{mod $2\pi$}, \\
		J_{n+1} & = J_n+K\sin\theta_n,
	\end{aligned}
\end{equation}
where $K>0$ is a constant.

Unlike the continuous-time Hamiltonian systems studied above, the standard map is an area-preserving map, so that it does not have any exit. However, we can construct exit basin diagrams by defining artificial leaks in the system, as explained in \cite{Sanjuan}. In particular, we define two leaks $L_1\equiv[(0.2-\omega)\pi,(0.2+\omega)\pi]\times[0,2\pi]$ and $L_2\equiv[(1.8-\omega)\pi,(1.8+\omega)\pi]\times[0,2\pi]$ (this choice guarantees that both leaks have width $\omega \pi$ and are symmetric about $\theta=\pi$). Thus, an exit basin is defined as the set of initial conditions falling after $1$ or more iterations in one particular leak. To represent exit basin diagrams, we simply assign a different color to the initial conditions depending on the first leak visited.

\begin{figure}[h!]
	\centering
	\includegraphics[clip,height=9cm,trim=0cm 0cm 1cm 0cm]{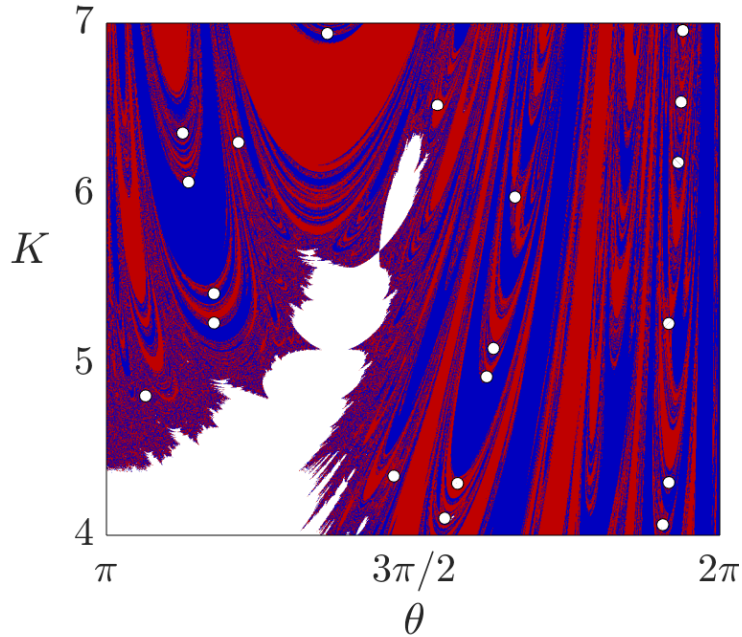}
	\caption{Islets of stability (solid white dots) in an exit basin diagram for the standard map with two symmetric leaks of width $0.1\pi$. The colors red and blue refer to initial conditions leading to the leaks $L_1$ and $L_2$, respectively. White regions inside the potential correspond to KAM islands. }
	\label{Fig10}
\end{figure}

For $K<4$, the periodic orbits of the system lie in the $\theta=0$ line, while for higher values of $K$ they appear in the lines $J=2\theta-2\pi$ and $J=2\theta$. We have searched for islets close to the value of $K$ where the main KAM island is destroyed, so we have computed exit basin diagrams in the $(\theta,K)$ plane following the line $J=2\theta-2\pi$ (we could have used the line $J=2\theta$ in an equivalent way). Therefore, once the value of $K$ and the initial condition $\theta_0$ are chosen, the initial condition in the $J$ coordinate is given by $J_0=2\theta_0-2\pi$. The result is shown in Fig.~\ref{Fig10}, where the position of $20$ islets is represented with white dots. The coordinate range where the islets can be found is shown in Table~\ref{T4} (see Appendix B).

  \section{Conclusions and discussion}\label{sec6}

In summary, our research reveals that the destruction of the main KAM island in two-degree-of-freedom Hamiltonian systems is explained by a cascade of period-doubling bifurcations. By using the Hénon-Heiles system as a model, we have calculated the conservative Feigenbaum constant and the accumulation point where the last periodic orbit becomes unstable. The value obtained for the Feigenbaum constant confirms that the geometrical progression of bifurcations is not only universal for area-preserving maps, but also for two-degree-of-freedom Hamiltonian systems.

We have also shown that not all KAM islands surround the main family of periodic orbits, but islets of stability exist for values above and below the accumulation point. We have studied these islets exhaustively, finding that all of them can be classified in three different types. The first type appears surrounding a stable periodic orbit created in a saddle-node bifurcation. The other two types emerge in the branches created in saddle-node bifurcations, always preceded by type I islets. To further demonstrate the validity of our classification scheme, we have identified the same types of islets in a different two-degree-of-freedom Hamiltonian system and in an area-preserving map.

We expect that this work could contribute to understand the formation, evolution, and destruction of KAM islands in Hamiltonian systems. The insights gained from this research may find applications in various physical systems where KAM islands play a critical role. Examples of such applications include plasma confinement in tokamaks \cite{Viana11}, chaotic transport of particles advected by fluid flows \cite{Solomon93}, and conductance fluctuations in chaotic cavities \cite{Ketzmerick96}.

 \section*{ACKNOWLEDGMENTS}
 This work has been financially supported by the Spanish State Research Agency (AEI) and the European Regional Development Fund (ERDF) under Project No. PID2019-105554GB-I00 (MCIN/AEI/10.13039/501100011033). 

\newpage

\section*{Appendix A: Propagation of uncertainty}
The energy values where period-doubling bifurcations occur have been calculated by detecting the change in the stability of periodic orbits. Our algorithm detects the values $E_s$ and $E_u$ for which the orbit is still sable and already unstable, respectively. Therefore, the bifurcation point is given by $E_n=(E_s+E_u)/2$ and its uncertainty by $\Delta E_n=(E_u-E_s)/2$. Since we use the $E_n$ for calculating $\delta_H$, its uncertainty is propagated as
\begin{equation*} 
	\begin{aligned}
		\Delta\delta_H & =\left|\frac{\partial \delta_H}{\partial E_n}\right|\Delta E_n+\left|\frac{\partial \delta_H}{\partial E_{n-1}}\right|\Delta E_{n-1}+\left|\frac{\partial \delta_H}{\partial E_{n-2}}\right|\Delta E_{n-2} \\
		& = \frac{(E_{n-1}-E_{n-2})\Delta E_n+(E_{n}-E_{n-2})\Delta E_{n-1}+(E_{n}-E_{n-1})\Delta E_{n-2}}{(E_{n}-E_{n-1})^2}. \\		
	\end{aligned}
\end{equation*}

In the case of the accumulation point $E_\infty$, its uncertainty is given by:

\begin{equation*} 
	\begin{aligned}
		\Delta E_{\infty} & =\left|\frac{\partial E_{\infty}}{\partial E_6}\right|\Delta E_6+\left|\frac{\partial E_{\infty}}{\partial E_7}\right|\Delta E_7+\left|\frac{\partial E_{\infty}}{\partial \delta_H}\right|\Delta \delta_H \\
		& = \frac{\Delta E_6+\delta_H\Delta E_7}{\delta_H-1}+\frac{(E_7-E_6)\Delta\delta_H}{(\delta_H-1)^2}. \\		
	\end{aligned}
\end{equation*}

\newpage
\section*{Appendix B: Coordinates of islets}

\begin{table}[h!]
	\begin{tabular}{c c c c c} 
		\hline
		$n$ \kern 2pc & $m$ & $E$ & $y$ & Type  \\ [0.5ex] 
		\hline\hline
		$1$ \kern 2pc&		$7$  & $[0.17668,0.17680]$ &  $[0.944,0.951]$ & I  \\ 
		\hline
		$2$ \kern 2pc &		$7$  & $[0.1768,0.17715]$ &  $[0.947,0.953]$ & II  \\ 
		\hline		
		$3$ \kern 2pc&$3$  & $[0.1838,0.1850]$ &  $[-0.372,-0.359]$ & I  \\ 
		\hline
		$4$ \kern 2pc&$3$  & $[0.1840,0.1865]$ &  $[-0.387,-0.372]$ & III  \\ 
		\hline
		$5$ \kern 2pc&$6$  & $[0.19188,0.19245]$ &  $[0.9496,0.9545]$ & III  \\
		\hline	
		$6$ \kern 2pc&	$6$  & $[0.19188,0.19245]$ &  $[-0.3845,-0.3810]$ & III  \\
		\hline
		$7$ \kern 2pc&	$7$  & $[0.193840,0.193865]$ &  $[0.5120,0.5165]$ & I  \\
		\hline	
		$8$ \kern 2pc&$7$  & $[0.199448,0.199464]$ &  $[0.13,0.136]$ & I  \\
		\hline	
		$9$ \kern 2pc& 	$7$  & $[0.20534,0.20552]$ &  $[0.2802,0.2835]$ & II  \\
		\hline
		$10$ \kern 2pc&	$5$  & $[0.20702,0.20716]$ &  $[0.504,0.511]$ & I \\
		\hline
		$11$ \kern 2pc&	$5$  & $[0.20735,0.20795]$ &  $[0.510,0.518]$ & II \\
		\hline	
		$12$ \kern 2pc&	$5$  & $[0.2123,0.2130]$ &  $[0.128,0.160]$ & I \\
		\hline
		$13$ \kern 2pc&	$5$  & $[0.2130,0.2143]$ &  $[0.122,0.136]$ & III  \\ 
		\hline
		$14$ \kern 2pc&	$3$  & $[0.217890,0.217903]$ &  $[0.1495,0.1530]$ & I  \\ 
		\hline
		$15$ \kern 2pc&	$3$  & $[0.21836,0.21853]$ &  $[0.1365,0.1390]$ & III  \\ 
		\hline
		$16$ \kern 2pc&	$3$  & $[0.2247,0.2252]$ &  $[0.675,0.703]$ & I  \\ 
		\hline
		$17$ \kern 2pc&    $3$ & $[0.2286,0.2308]$ & $[0.728,0.0.740]$ & III \\
		\hline
		$18$ \kern 2pc&	$5$  & $[0.23405,0.23437]$ & $[0.9767,0.9787]$ & II \\  	
		\hline
		$19$ \kern 2pc&   $5$ & $[0.23623,0.23630]$ & $[-0.2212,-0.2185]$ & I \\
		\hline
		$20$\kern 2pc &	$5$ & $[0.23627,0.23643]$ & $[-0.2185,-0.2157]$ &III \\
		\hline
		$21$ \kern 2pc&	$1$ & $[0.25285,0.25345]$ & $[-0.11,-0.07]$ & I \\
		\hline
		$22$ \kern 2pc&	$3$ & $[0.260769,0.260782]$ & $[-0.1005,-0.095]$ & I \\  
		\hline
		$23$ \kern 2pc&	$3$ & $[0.260769,0.260782]$ & $[0.98557,0.98600]$ & I \\  
		\hline
		$24$ \kern 2pc&	$1$  & $[0.26194335,0.26194367]$ & $[-0.0993,-0.0984]$ & I \\  
		\hline		
	\end{tabular}
	\caption{Range of coordinates in the $(y,E)$ plane of the Hénon-Heiles system where several islets of stability of different multiplicity and type can be found.}
	\label{T2}
\end{table}

\begin{table}
	\begin{tabular}{c c c c c} 
		\hline
		$n$ \kern 2pc & $m$ & $E$ & $x$ & Type  \\ [0.5ex] 
		\hline\hline
		$1$ \kern 2pc&		$7$  & $[0.330768,0.330815]$ &  $[-0.56056,-0.56048]$ & III  \\ 
		\hline
		$2$ \kern 2pc &		$7$  & $[0.35304,0.35311]$ &  $[-0.57842,-0.57826]$ & III  \\ 
		\hline	
		$3$ \kern 2pc&$11$  & $[0.387718,0.387727]$ &  $[-0.55975,-0.55965]$ & II  \\ 
		\hline
		$4$ \kern 2pc&$3$  & $[0.572922,0.572932]$ &  $[-0.9473,-0.9458]$ & I  \\ 
		\hline
		$5$ \kern 2pc&$13$  & $[0.357048,0.357049]$ &  $[-0.774675,-0.774620]$ & I  \\
		\hline	
		$6$ \kern 2pc&	$13$  & $[0.3570487,0.3570515]$ &  $[-0.774735,-0.774695]$ & III  \\
		\hline
		$7$ \kern 2pc&	$5$  & $[0.377254,0.377257]$ &  $[-0.7471,-0.7463]$ & I  \\
		\hline	
		$8$ \kern 2pc&$7$  & $[0.374565,0.374595]$ &  $[-0.8487,-0.8482]$ & I  \\
		\hline	
		$9$ \kern 2pc& 	$7$  & $[0.375200,0.375455]$ &  $[-0.84763,-0.84755]$ & III  \\
		\hline
		$10$ \kern 2pc&	$7$  & $[0.471438,0.471448]$ &  $[-0.66935,-0.66885]$ & I \\
		\hline
		$11$ \kern 2pc&	$7$  & $ [0.47153,0.47160]$ &  $[-0.6683,-0.6680]$ & III \\
		\hline	
		$12$ \kern 2pc&	$1$  & $[0.21330,0.21355]$ &  $[0.093,0.113]$ & I \\
		\hline	
	\end{tabular}
	\caption{Range of coordinates in the $(x,E)$ plane of the Barbanis system where several islets of stability of different multiplicity and type can be found.}
	\label{T3}
\end{table}

\begin{table}
	\begin{tabular}{c c c c} 
		\hline
		
		$n$ \kern 2pc & $K$ & $\theta$ & Type  \\ [0.5ex] 
		\hline\hline
		$1$ \kern 2pc&	 $[5.985,5.996]$ &  $	[4.061,4.066]$ & III  \\ 
		\hline
		$2$ \kern 2pc &	 $[4.85,4.92]$ &  $	[4.07,4.12]$ & I  \\ 
		\hline	
		$3$ \kern 2pc&	 $	[6.020,6.026]$ &  $[4.309,4.313]$ & II  \\ 
		\hline
		$4$ \kern 2pc&	 $	[4.936,4.942]$ &  $[4.3036,4.3042]$ & I  \\ 
		\hline
		$5$ \kern 2pc&	 $	[4.608,4.616]$ &  $[4.346,4.362]$ & III  \\
		\hline	
		$6$ \kern 2pc&	 $[3.339,3.343]$ &  $[4.816,4.819]$ &II  \\
		\hline
		$7$ \kern 2pc&	 $[5.081,5.091]$ &  $[4.927,4.930]$ & I  \\
		\hline	
		$8$ \kern 2pc  &	 $[5.115,5.135]$ &  $[5.090,5.097]$ & I  \\
		\hline	
		$9$ \kern 2pc  &	 $	[6.013,6.030]$ &  $	[5.235,5.265]$ & III  \\
		\hline
		$10$ \kern 2pc  &	 $[3.684,3.694]$ &  $[5.2431,5.2434]$ & I \\
		\hline
		$11$ \kern 2pc  &	 $[3.687,3.692]$ &  $[5.414,5.417]$ &III \\
		\hline	
		$12$ \kern 2pc  &	 $[5.21,5.25]$ &  $	[5.95,6.00]$ & III \\
		\hline	
		$13$ \kern 2pc  &	 $	[3.557,3.573]$ &  $[6.067,6.070]$& I \\
		\hline	
		$14$ \kern 2pc  &	$[6.065,6.071]$ &  $[6.180,6.185]$ & III \\
		\hline	
		$15$ \kern 2pc  &	 $[3.813,3.819]$ &  $[6.296,6.304]$ & II \\
		\hline	
		$16$ \kern 2pc  &	 $[3.521,3.538]$ &  $[6.352,6.360]$& II \\
		\hline	
		$17$ \kern 2pc  &	 $[4.832,4.834]$ &  $[6.521,6.523]$ & III \\
		\hline	
		$18$ \kern 2pc  &	$[6.083,6.085]$ &  $[6.542,6.543]$ & II \\
		\hline	
		$19$ \kern 2pc &	$[4.24,4.32]$ &  $[6.935,6.945]$ &I \\
		\hline	
		$20$ \kern 2pc  &	 $[6.091,6.093]$ &  $[6.9568,6.9572]$ & I \\
		\hline	
	\end{tabular}
	\caption{Range of coordinates in the $(\theta,K)$ plane of the standard map where several islets of stability can be found.}
	\label{T4}
\end{table}

\end{document}